\documentclass[aps,prl,reprint,superscriptaddress,showpacs]{revtex4-1}

\usepackage{graphics,graphicx,epsfig}
\usepackage{amssymb,color}
\usepackage{epsf,epstopdf,wrapfig}
\usepackage{mciteplus}
\usepackage{graphicx,color}
\usepackage{amsmath,amssymb}
\usepackage{dcolumn}   
\usepackage{bm}        
\usepackage{multirow}

\def\be{\begin{equation}}
\def\ee{\end{equation}}
\def\bea{\begin{eqnarray}}
\def\eea{\end{eqnarray}}

\def\ra{\rangle}
\def\la{\langle}

\newcommand{\eref}[1]{Eq.~(\ref{#1})}%
\newcommand{\fref}[1]{Fig.~\ref{#1}} %
\newcommand{\aref}[1]{\hyperref[#1]{Appendix~\ref{#1}}}

\begin{document}

\title{Role of spatial heterogeneity on the collective dynamics of cilia beating in a minimal 1D model}

\author{Supravat Dey}
\email{supravat.dey@gmail.com}

\affiliation{L2C, Univ Montpellier, CNRS, Montpellier, France.}
\thanks{\emph{Present address of SD:} School of Physics and Astronomy, Rochester Institute of Technology, Rochester, New York 14623, USA}

\author{Gladys Massiera}
\email{gladys.massiera@umontpellier.fr}
\affiliation{L2C, Univ Montpellier, CNRS, Montpellier, France.}
\author{Estelle Pitard}
\email{estelle.pitard@umontpellier.fr}
\affiliation{L2C, Univ Montpellier, CNRS, Montpellier, France.}


\pacs{87.16.Qp,05.45.Xt,47.63.-b}

\begin{abstract}

Cilia are elastic hairlike protuberances of the cell membrane found in various unicellular organisms and in several tissues of most living organisms. In some tissues such as the airway tissues of the lung, the coordinated beating of cilia induce a fluid flow of crucial importance as it allows the continuous cleaning of our bronchia, known as mucociliary clearance. While most of the models addressing the question of collective dynamics and metachronal wave consider homogeneous carpets of cilia, experimental observations rather show that cilia clusters are heterogeneously distributed over the tissue surface. The purpose of this paper is to investigate the role of spatial heterogeneity on the coherent beating of cilia using a very simple one dimensional model for cilia known as the rower model. We systematically study systems consisting of a few rowers to hundreds of rowers and we investigate the conditions for the emergence of collective beating. When considering a small number of rowers, a phase drift occurs, hence a bifurcation in beating frequency is observed as the distance between rowers clusters is changed. In the case of many rowers, a distribution of frequencies is observed. We found in particular the pattern of the patchy structure that shows the best robustness in collective beating behavior, as the density of cilia is varied over a wide range. 

\end{abstract}

\maketitle

\section{I. Introduction}

Cilia are elastic hairlike protuberances of the cell membrane found in various unicellular organisms and in several tissues of most living organisms. As a propulsor, a cilium is periodically beating in a succession of power and recovery strokes, propelled by its internal molecular motors \cite{Graybook}. Propulsion acts either on the microorganism itself such as for {\it Paramecium} or {\it Volvox} algae, or to induce the flow of the surrounding fluid, as this the case for airway, brain or oviduct tissues \cite{Ines2003,Sawamoto2006,SmithRes08}. In the airway tissues of the lung, this fluid flow induced by the coordinated beating of cilia is of crucial importance as it allows the biological function of cilia to help to expel the mucus and the impurities out of the airways, known as mucociliary clearance \cite{SmithRes08}.

Cilia are often observed as scattered clumps and beat in coordinated manner in the form of metachronal waves keeping a constant phase difference with adjacent cilia \cite{Brennen1977,Sanderson81,Brumley2012}. This type of large scale coordinated beating pattern is of great importance for efficient propulsion \cite{Gueron1999,Osterman2011,Elgeti2013,Michelin2010,Goldstein2016}. It is now well established that hydrodynamic interactions play a crucial role \cite{GolestanianSM2011,Brumley2012,Elgeti2013,Bruot2016} for the emergence of such large scale metachronal waves.

While most of the models addressing the question of collective dynamics and metachronal wave consider homogeneous carpets of cilia, in real samples, cilia form patches. In cultured and in vivo airways epithelium tissues, cilia patches are heterogeneously distributed over the surface \cite{Gabridge77, HoffmannJMed14}. For example, observations performed on human bronchial epithelium cultures show that cilia are distributed in clusters containing $\sim 100-200$ cilia \cite{Blake1975}, and that these clusters are separated in a random manner. The mucociliary dysfunction due to impaired coordinated beating of cilia is poorly understood and could be related to the spatial heterogeneity of healthy cilia distribution. The main focus of this work is to study the collective behavior of a system of many hydrodynamically coupled beating cilia with heterogeneous spatial configurations.

Several theoretical models with various levels of complexity investigate the relation between the hydrodynamic coupling and the metachronal synchronization \cite{GolestanianRev11,Bruot2016}. We will concentrate our study on  a minimal model, where the beating pattern is simple and is composed of a few degrees of freedoms allowing us to understand the sole role of hydrodynamical coupling in the beating synchronization of an array of cilia. There are two classes of well studied minimal models. In one class of models, a cilium is described as a rower: a spherical bead oscillates  between two distinct states where, in each state, the bead moves in a specific driving potential, mimicking the power and recovery strokes of a real cilium  \cite{CosentinoPre03,CosentinoSM09,CosentinoPRE10,CicutaPnas10,CosentinoPRL11,CosentinoSM12,CosentinoPRE12,Bruot2016}. In another class models, cilia (known as rotors) are considered as spherical beads orbiting on rigid or flexible two dimensional trajectories under a driving torque  \cite{Niedermayer2008, UchidaPrl10,GolestanianSM2011,UchidaPrl2011,GolestanianRev11}. In more  complex models, cilia are modelled as actively driven semiflexible filaments with more realistic  beating pattern \cite{GueronPnas97,JoannyBioPhys07,Elgeti2013}. In such models, the hydrodynamic coupling between cilia can lead to metachronal waves in systems of cilia in one- and two-dimensional lattices. In this paper, to understand the role of spatial heterogeneity, we consider the framework of the rower model.

In this work, we will present new results on how heterogeneity of cilia position influences the stability and robustness of coherent beating of cilia. We focus on one of the simplest models for cilia beating, the rower model \cite{CosentinoPre03} in one dimension. We carefully study the effect of spatial heterogeneity in systems with few rowers as well as in systems with a large number of rowers with various kinds of heterogeneities. First, we study the 3-rowers case in great detail as it is the minimal way to introduce  position irregularities. We find phase drifting and bifurcation of the frequencies of beating when the distance between the second and third rower is varied, keeping the distance between the first and second rowers fixed. The phenomenon of phase drifting is also present in a system with more than 3 rowers, and we study different spatial configurations with a finite number of rowers. This allows us to identify a crossover distance, above which the separation between consecutive clusters is large enough to decouple their dynamical behavior. In the case of a large number of coupled rowers, we consider several types of spatial heterogeneity. We find that when the cilia are spatially distributed on randomly clustered configurations, their dynamical behaviour is characterized by a robust average common beating frequency, that depends only weakly on the density of cilia. This seems to correspond to preliminary results on human epithelial tissues \cite{Kamel2015}. On the other hand, the collective behaviour observed for other types of  spatial heterogeneities is strikingly different.

\section{II. The Rower model}
\begin{figure*}[]
\centering
\mbox{
\includegraphics[scale=0.4]{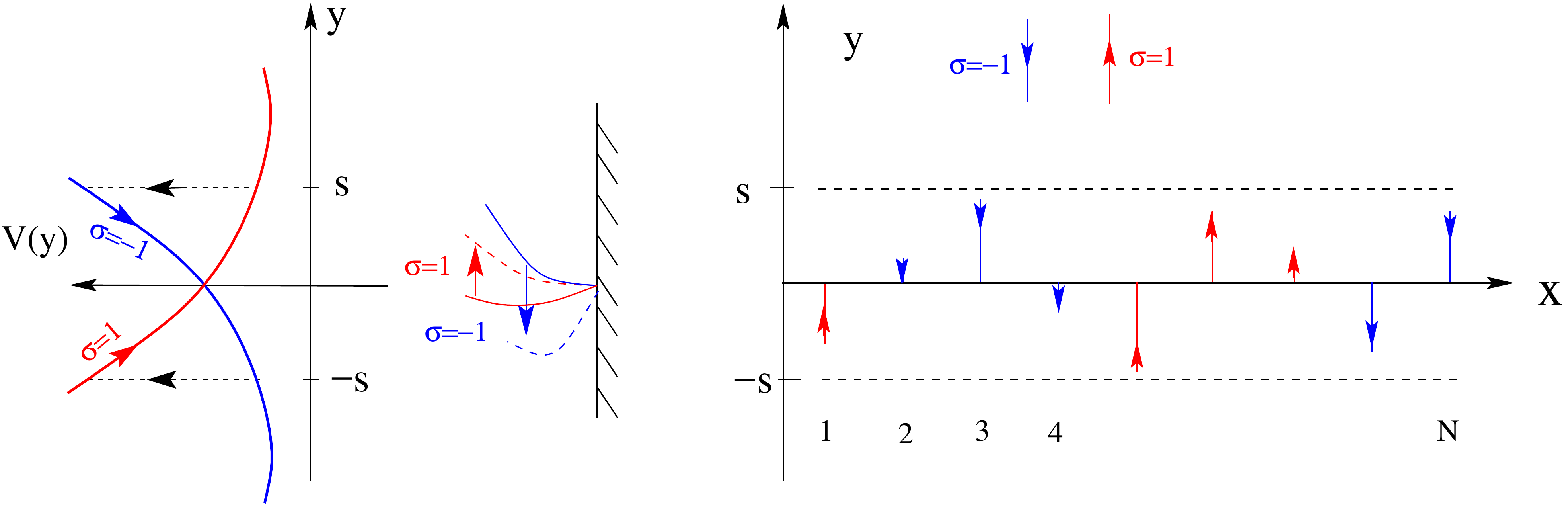} 
}
\caption{A schematic diagram of the model. {\bf Left} The harmonic potentials correspond to the two different states ($\sigma=\pm1$) of a rower. The bead switches between the two potentials once it reaches $y=\pm s$. The driving potentials and the switching mechanism form a simple description of real beating. {\bf Middle} A schematic representation of simplified beating pattern of a cilium. {\bf Right} $N$ rowers are placed on a regular one dimensional lattice. The positive $y$ direction is $\sigma=1$ and negative $y$ direction is $\sigma=-1$. We create heterogeneity in rowers' position by keeping empty lattice sites in-between occupied sites. These are shown in \fref{fig:fig11}.}
\label{fig:fig1}
\end{figure*}

The rower model for cilia,  proposed by Cosentino {\it et al.} in 2003 \cite{CosentinoPre03}, experimentally realized by driven colloids in viscous fluids \cite{CicutaPnas10}, remains one of the basic model to study hydrodynamic synchronization \cite{Bruot2016}. In this model, the complex structure of a cilium is coarse-grained as a spherical bead, and the periodic beating pattern is described as  an oscillating linear motion of a bead in a viscous fluid. In order to create a sustained oscillating motion of the bead, two different driving potentials are used, and a mechanism for geometrical switching between these potentials is employed. In \fref{fig:fig1} (left), we show a schematic diagram of the geometric switch and the harmonic potentials. The bead motion under each driving potential corresponds to a specific state of the bead $\sigma=\pm 1$ (motion corresponding to $\pm y$ direction). If the bead reaches a particular limiting amplitude $y = \pm s$, it switches to the other driving potential and consequently  reverses the direction of its motion. Hence, it successfully creates a sustained oscillating motion. These two states of the bead mimic the power and recovery stroke of beating of cilia. 

We consider an array of $N$ rowers on a one dimensional lattice of size $L$ (see \fref{fig:fig1} (right)). The spacing between two consecutive lattice sites is $d=1$. If there is no heterogeneity, all the lattice sites are occupied and hence, $L=N$. In the case where rowers are placed heterogeneously on the lattice, one allows for empty lattice sites and $L>N$. Let us call the positions of the rowers  $x_1, x_2, .., x_n, .., x_N$ where $x_1<x_2, ..,<x_n, .., <x_N$. Two dynamical variables $\sigma_n$ (state) and $y_n$ (displacement) describe the motion of the $n^{th}$ rower.

A bead switches between the two states if the displacement $y$ reaches the maximum amplitude $\pm s$, i.e. if $y_n(t)=\pm s$ then $\sigma_n(t)=-\sigma_n(t)$ and $\sigma_n(t)=\sigma_n(t)$ otherwise. We choose the two driving potentials to be harmonic  $V(y_n,\sigma_n)$ (\fref{fig:fig1} (left)). The external driving force $f_n$ for the $n^{th}$ rower is given by:
\bea 
f_n =  -\frac{{\partial V}(y_n, \sigma_n)}{\partial y_n} = -(k y_n -\sigma_n).
\eea
In our study, the stiffness constant of the potential $k$ is assumed to the same for all the rowers. This external driving force on the bead is a simple approximation of the complex internal active force of a real cilium.

Cilia motion corresponds to a low Reynolds number regime and we are interested by  the far-field hydrodynamic regime: the size and displacements of the beads are  small compared to the lattice spacing. In this limit, the hydrodynamic coupling between the rowers is given by Oseen tensor \cite{Happel83, Dhont96}. The velocity at any instant of time $v_{mn}$ acting on the $n^{th}$ rower induced by active oscillation of rower $m$ is given by:
\bea
v_{mn} = O(m,n) f_m.
\eea
The Oseen coupling between any two rowers $m$ and $n$ is $O(m,n)=1/(8\pi\eta d_{mn})$, where $d_{mn}$ is the distance between the sites ($d_{nm}=|x_n-x_m|$), and $\eta$ is the viscosity of the fluid medium. Here, we note that as our focus in this paper is on the coherence of cilia beating (not on the flow of surrounding fluid) and therefore the driving forces for both states are chosen to be symmetric for simplicity (as in \cite{CicutaPnas10,CosentinoSM12,CosentinoPRE12}). However, one can make the motion of the two states asymmetric by considering different driving forces \cite{StarkEpj11} or different drag coefficients for both states \cite{CosentinoPre03}.

Hence, in the overdamped limit, the dynamical equations for this system are given by,
\bea
\nonumber
\dot{y}_n &=& \frac{1}{6 \pi \eta a}f_n + \sum_{m\neq n}^Nv_{mn}\\ 
           &=& \frac{1}{6 \pi \eta a}f_n + \sum_{m\neq n}^N \frac{f_m}{8 \pi \eta d_{mn}}.
\label{eqn:dynamics0}
\eea
Here, the factor $6\pi\eta a$ is the viscous drag coefficient of a spherical bead with radius $a$. For computation purposes, we choose $6 \pi \eta a = 1$, hence $1/(8 \pi \eta)= 3 a/4 \equiv \alpha$. Replacing $a$ and $\eta$, \eref{eqn:dynamics0} can be rewritten as,
\bea
\dot{y}_n &=& f_n + \alpha \sum_{m\neq n}^N \frac{f_m}{d_{mn}}.
\label{eqn:dynamics}
\eea
Note that in a realistic physical situation, the coupling strength $\alpha$ is always a positive quantity. In the absence of any hydrodynamic coupling, (i.e. $\alpha=0$), it is easy to calculate the natural frequency of the rowers $\omega_0$. It depends on the force constant $k$ and the value of limiting displacement $s$; its analytical expression is given by \cite{CosentinoPre03,CicutaPnas10}:
\bea
\omega_0 = \frac{2 \pi}{T} = \frac{\pi k}{\ln\left[\frac{1+ks}{1-ks}\right]}.
\label{eqn:omega_0}
\eea

As mentioned before, this model has been studied for systems of several rowers on regular lattices in one and two dimensions \cite{CosentinoPre03, StarkEpj11}. The collective beating of the rowers lead to metachronal waves as a result of anti-phase synchronization of neighbours for $k>0$. When $k<0$ \cite{StarkEpj11}, or $\alpha<0$ \cite{CosentinoPre03} (which is not realistic) the rowers show in-phase oscillations, not metachronal waves. Cases considering a few rowers in special geometries were also studied \cite{CosentinoSM12,CosentinoPRE12,RoyCBM13} and showed very different dynamical states depending on their spatial configuration and orientation of beating. However, no previous study have considered the spatial heterogeneity of the 
cilia position \cite{RoyCBM13}.

For our computation, we choose  the values for the parameters $\alpha$, $k$ and $s$ already used in the original paper \cite{CosentinoPre03}. The force constant for the harmonic driving potential is $k=1.0$. A single rower oscillates between $-s$ to $+s$ with $s=0.8$. The value of $\alpha$ is taken to be $0.1$ unless otherwise mentioned. In order to integrate the dynamical equations (\eref{eqn:dynamics}) for $N$ rowers, we use the Euler method with an integration step $h$ between $10^{-5}$ and $10^{-3}$, depending on the situation. We consider here the  deterministic case, hence simulations are run without thermal noise. The case of adding thermal noise will be considered elsewhere \cite{Dey_inprep}.

In the following, we try to understand the role of spatial heterogeneity on the collective behavior of cilia beating. First, we study a 3-rower system which is the minimal sets to study heterogeneity effect. We consider 3 rowers on a one dimensional lattice  such that the two first rowers occupy two consecutive lattice sites  whereas the position of the  third rower increases (see \fref{fig:fig4}). Then we study the effect of heterogeneity in systems composed of  4, 5, and 10 rowers, where we find that the resulting dynamics obeys a general scenario (section III). Finally, we study systems of a large number of rowers with three types of heterogeneities (section IV)corresponding to different degrees of randomness --- (i) the regular clustered case, (ii) random case, and (iii) the random clustered case.

\subsection{How to characterize collective behavior?}

A collection of oscillators can display different emerging features through coherence in phases and frequencies \cite{Pikovsky03,JuanRevMod05}. Most common phenomena of phase coherence are synchronization and phase-locking. In order to characterize the phase coherence, we define a phase variable $\phi_n$ for $n^{th}$ rower using the following prescription (given by Stark {\it et al.} \cite{StarkEpj11}):
\bea
\nonumber
\phi_n &=& 2 \pi m_n + \frac{\pi}{2} \sigma_n \frac{y_n}{s}, ~ \rm{if~y_n>0~and~\sigma_n=1},\\
\nonumber
       &=& 2 \pi m_n + \frac{\pi}{2} \sigma_n \frac{y_n}{s} + 2 \pi, ~\rm{if~y_n<0~and~\sigma_n=1},\\
       &=& 2 \pi m_n + \frac{\pi}{2} \sigma_n \frac{y_n}{s} + \pi, ~\rm{if~\sigma_n=-1.}
\label{eqn:phaserepresentation}
\eea
\begin{figure}[h]
\centering
\includegraphics[scale=0.3,angle=-90]{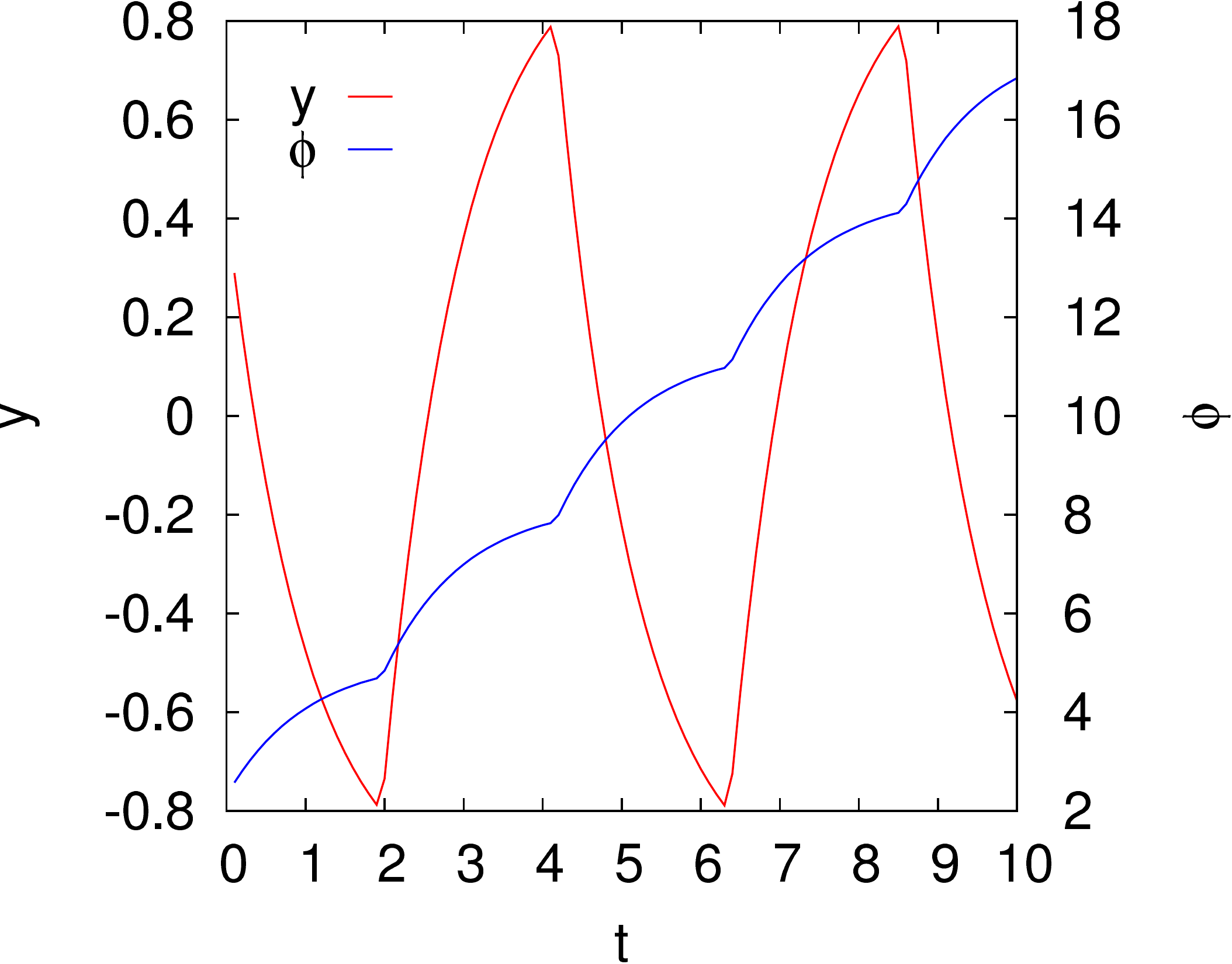}
\caption{The periodic beating of a single rower is plotted as function of time $t$. The red curve is the displacement $y$ of the rower, and the blue curve is its phase representation according to \eref{eqn:phaserepresentation}.}
\label{fig:fig2}
\end{figure}
The phase number  $m_n$ ($\in \mathbb{N}$) is increased by $1$ after rower $n$ completes a full cycle. The piecewise linear relation between phase and displacement ensures that one full cycle in $y$ is equivalent to a rotation of $2 \pi$ in $\phi$. In \fref{fig:fig2}, we plot the time evolution of displacement $y$ and corresponding phase $\phi$ for a single rower. In the case of perfect synchronization, all oscillators oscillate in  phase;  it is defined as $\phi_j=\phi_{j+1}$ for all $j$. For phase-locking systems, neighbouring oscillators maintain a constant nonzero phase difference $\delta$,  $\phi_j=\phi_{j+1} + \delta$. A nonzero $\delta$ leads to the formation of traveling wave in a system of many oscillators. This wave is known as metachronal wave in the literature \cite{Sanderson81}. The degree of phase coherence can be measured by a complex order parameter $Z$, which is defined as
\cite{StarkEpj11}, 
\bea
Z = A {\rm e}^{i \Phi} = \frac{1}{N-1} \sum_{n=1}^{N-1} \rm{e}^{i \, \Delta \phi_n},
\label{eqn:orderpara}
\eea
where $\Delta \phi_n = \phi_{n+1}  - \phi_n$ and $N$ is the total number of oscillators. The system is maximally coherent when $A=1$, and has no coherence for $A=0$ (see Appendix \ref{app:orderpara} I). When $\Delta \phi_n=\delta$ for all $n$, $A=1$. This is true for any constant $\delta$ including zero i.e, for both phase locked and in phase synchronization solutions. Therefore, for both perfect metachronal wave and fully synchronized states the value of the order parameter $A$ is 1.  In order to distinguish the synchronized states from metachronal waves, one should also measure $\Phi$, which gives the average angle of phase difference. For a synchronized state $\Phi=0$, and for a metachronal wave, $\Phi=\delta\neq0$.

Beside phase coherence, another observed  emerging behavior in a system of coupled oscillators is {frequency locking}. In this case, the phases of oscillators can be different but they oscillate with a common frequency which is different from their natural frequency (the frequency in the absence of coupling) \cite{Pikovsky03, JuanRevMod05}. We define the average frequency ${\omega_i}$ of a rower $i$ in the following way \cite{JuanRevMod05},
\bea
\omega_i = \lim_{t\to\infty} \frac{1}{t} \int_{t_0}^{t_0+t}\dot{\phi}_i\,dt 
= \lim_{t\to\infty}\frac{\phi_i(t_0+t)-\phi_i(t_0)}{t}.
\label{eqn:defomega}
\eea
Here, $t_0$ is sufficiently large so that the system has reached a dynamical steady state at that time. In general, the  frequencies of the oscillators are distributed over a distribution $P(\omega)$. For a perfectly frequency locked system, the probability distribution of frequencies $P(\omega)$ is a $\delta$-function.

\section{III. Results for finite-size systems}

In this section, we systematically study arrays consisting of a few rowers (from $N=2$) to many rowers ($N=100$). We compute the order parameter $A$ (\eref{eqn:orderpara}), and the distribution of beating frequencies $\omega_i$ (\eref{eqn:defomega}) to characterize the collective behavior of the beating.  Our  objective is to understand the role of the spatial heterogeneity of cilia on the stability and robustness of the synchronization. This heterogeneity is introduced as soon as the number of rowers is 3.

\subsection{1. 2-rowers case }

The dynamical behavior of a system two rowers are well understood. It is known that two rowers oscillate in opposite phase \cite{CosentinoPre03,StarkEpj11,CicutaPnas10}, and the collective frequency of the oscillation depends on the separation and hydrodynamic strength \cite{CicutaPnas10}. Here, we revisit the two-rower case mainly to study collective frequency for different separating distances $d_{12}$. As we will see in the next sections, this study will be useful to understand a system with more rowers.

\begin{figure}[h]
\mbox{
\hspace{-1cm}
\includegraphics[angle=-90,scale=0.25]{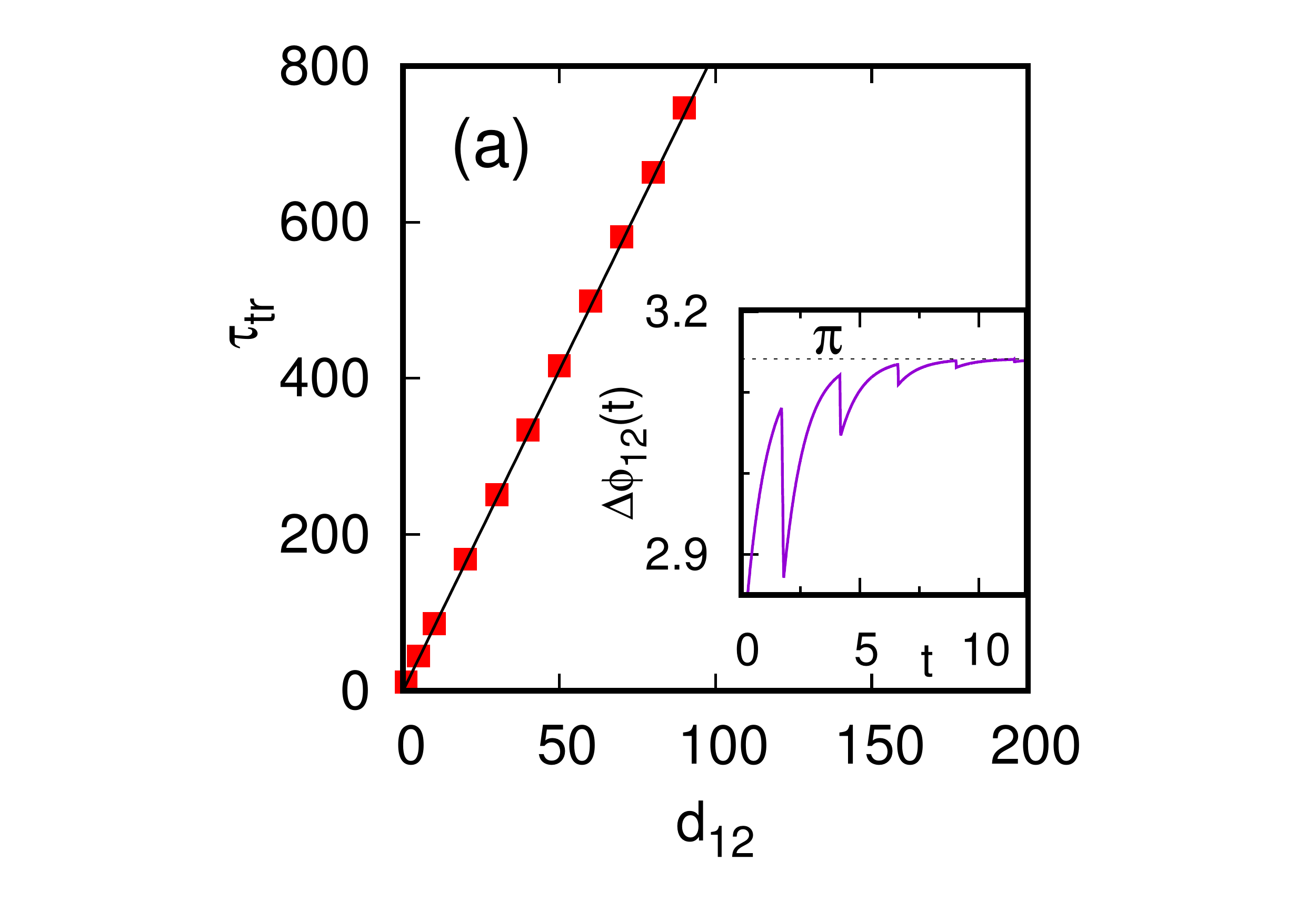}
\hspace{-2.3cm}
\includegraphics[scale = 0.25,angle=270]{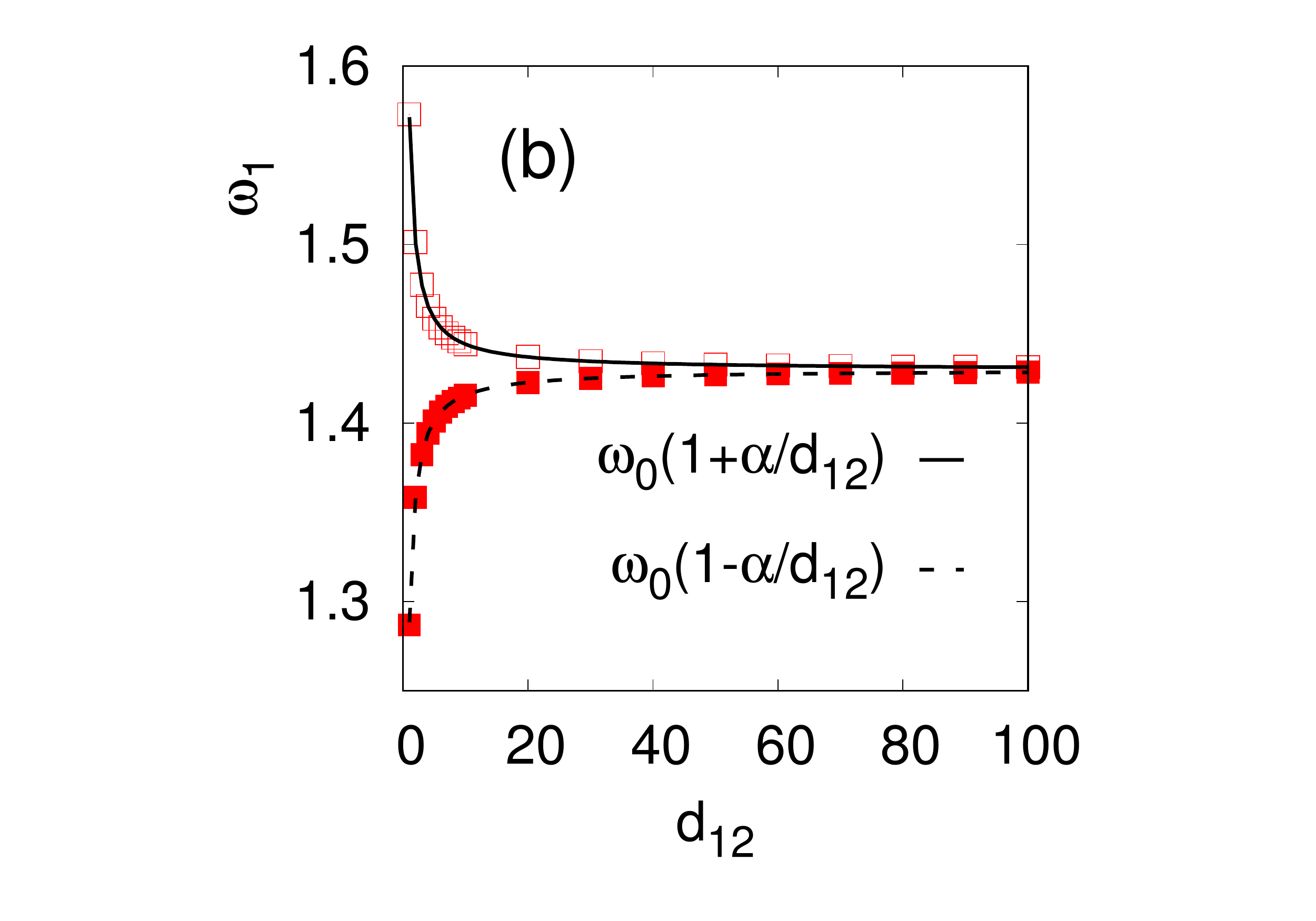}
}
\caption{(a) The transient time ($\tau_{tr}$) to reach the anti-synchronized state from an arbitrary initial condition is plotted against separation $d_{12}$. $\tau_{tr}$ increases linearly with $d_{12}$. Inset: The phase difference between the rowers, $\Delta\phi_{12}$, is plotted against time $t$. At $t=0$, the rowers start from an arbitrary initial condition, and after $\tau_{tr}$, $\Delta\phi_{12}$ converges to $\pi$, i.e. the rowers beat in exact anti-phase synchronization. (b) Collective frequency $\omega_{coll,2}(d_{12}$) as a function of $d_{12}$ for different kinds of initial conditions. The data with solid squares are obtained when the two rowers start from a initial condition which is different from an exact in-phase configuration while the data with empty squares are obtained with in exact in-phase initial condition. Their frequencies converge to $\omega_0$ for large $d_{12}$.}  
\label{fig:fig3}
\end{figure}

In the inset of \fref{fig:fig3}(a), the phase difference $\Delta\phi_{12}(t)=\phi_2(t)-\phi_1(t)$ is plotted as a function of time $t$ for $\alpha=0.1$. At $t=0$, we start the simulation with an arbitrary phase difference. We observe that within a small transient time the 2 rowers reach to a perfect anti-phase synchronization state ($\Delta\phi_{12}(t)=\pi$). Let $\tau_{tr}$ be the average time required to reach a perfect anti-phase synchronized state. In our simulation, we compute $\tau_{tr}$ by the time $t$ at which $\Delta\phi_{12}(t)$ reaches the value $0.99\pi$ and then averaging the data over many random initial configurations($\sim 1000$). We find that $\tau_{tr}$ linearly increases with $d_{12}$ when $\alpha$ is constant (or with $1/\alpha$ when $d_{12}$ is constant) (see \fref{fig:fig3}(a)). This is consistent with the fact that the larger the interaction ($\alpha/d_{12}$), the faster the rowers reach the synchronization state.

What is the collective frequency of the anti-phase synchronization for two rowers? One can derive the expression for the collective frequency $\omega_{coll,2}(d_{12})$, by assuming the anti-phase solution in the dynamical equation \eref{eqn:dynamics} \cite{CicutaPnas10}. The collective frequency $\omega_{coll,2}(d_{12})$  is a function of $d_{12}$ and $\alpha$, and is given by:
\be
\omega_{coll,2}(d_{12}) = \omega_{1,2} = \omega_0 (1-|\alpha|/d_{12}),
\label{eqn:collectivefreq2}
\ee
where $\omega_0$ is the natural frequency of the rower when the interaction is absent (see \eref{eqn:omega_0}). This matches exactly with the numerical findings, as shown in \fref{fig:fig3}(b).

A special case arises for a specific initial condition, which is specific to the deterministic (zero noise) system. At $t=0$, if the rowers are in exact in-phase ($y_1=y_2$ and $\sigma_1=\sigma_2$) there will not be any anti-phase synchronization, and the rowers will continue to beat in phase forever.
The analytical expression for $\omega_{coll,2}$ of the beating is given by:
\be
\omega_{coll,2}(d_{12}) = \omega_{1,2} = \omega_0 (1+|\alpha|/d_{12}).
\label{eqn:collectivefreq2_special}
\ee
In  \fref{fig:fig3}(b), we plot the frequency as a function of $d_{12}$ when the rowers start from an exact in-phase initial condition (empty squares): the data points exactly match  the analytical expression given by \eref{eqn:collectivefreq2_special}. However, this solution is very unstable. If the simulation starts with an initial condition which is slightly different from those specific initial conditions,  one reaches  the usual solution discussed above.

\subsection{2. 3-rowers case}

A 3-rowers system is the minimal setup to study the spatial heterogeneity  of cilia. It shows an interesting phenomenon called phase drifting in which the third rower oscillates with a different phase than the first two rowers, typical of a dynamical bifurcation. This phenomenon is familiar in other coupled non-linear systems of oscillators when the coupling strength or the noise strength is varied (Adler systems \cite{Adler46,Stratonovich63,Pikovsky03, Friedrich2016}), and has also been reported for rotors near a wall as the distance from the wall is increased \cite{Brumley2016}.

\begin{figure}[h]
\centering
\includegraphics[scale =0.5]{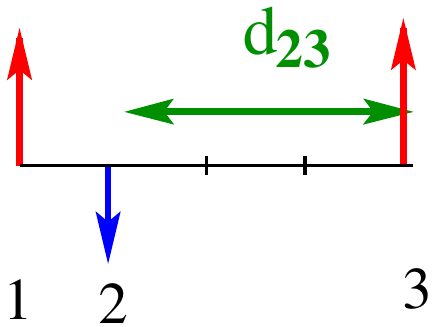}
\caption{The minimal set up to study the spatial inhomogeneity of cilia position. The first two rowers occupy the two consecutive lattice sites $d_{12}=1$. The third is placed $d_{23}$ distance apart from the second rower.}
\label{fig:fig4}
\end{figure}

We consider the 3-rowers system in one dimension, schematically shown in  \fref{fig:fig4}, where  the first two rowers occupy two consecutive lattice sites and the third rower is placed on a site which is a distance $d_{23}$ apart from the second rower. The lattice constant is $d=d_{12}=1$. We consider different cases for different values of $d_{23}$.  

\begin{figure}[]
\centering
\mbox{
\hspace{-1cm}
\includegraphics[scale = 0.235,angle=270]{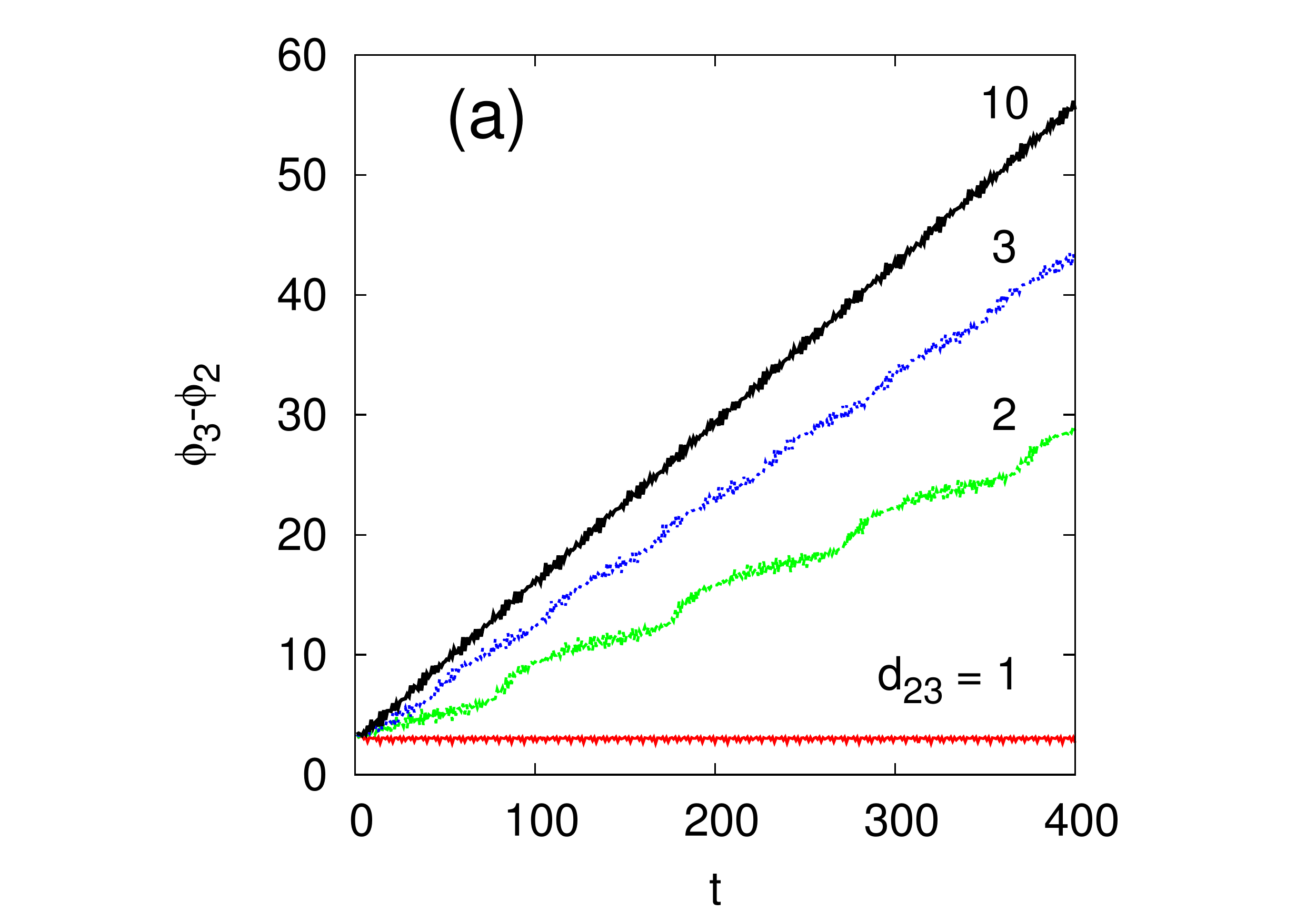}
\hspace{-2cm}
\includegraphics[scale = 0.255,angle=270]{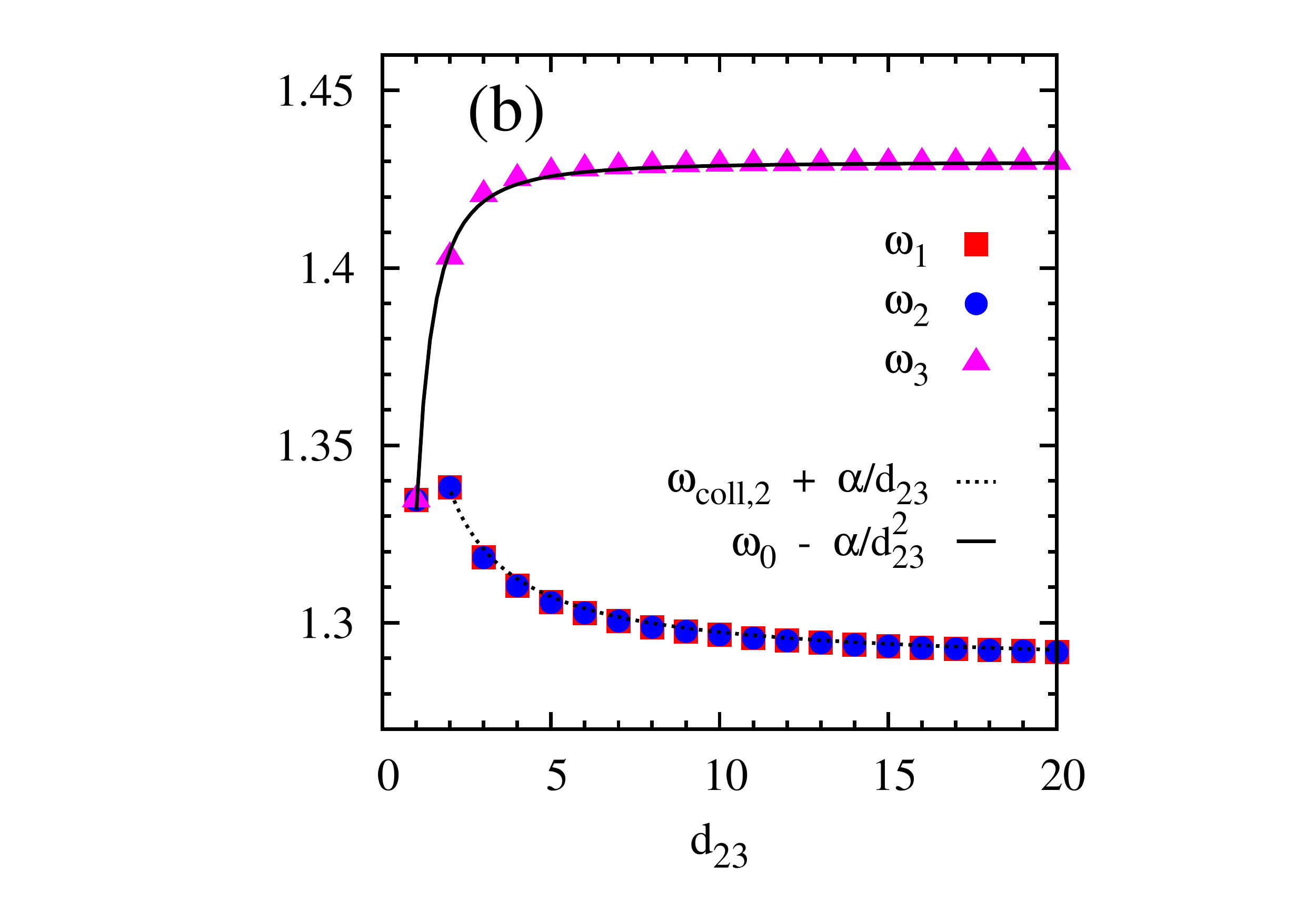}
}
\caption{(a) The phase difference between the rower 2 and 3, $\Delta\phi_{23}$ versus time $t$ for different $d_{23}$. $\Delta\phi_{23}$ remains constant for $d_{23} < d_c$.  At $d_c$, a bifurcation occurs and  $\Delta\phi_{23}$ grows in time for $d_{23}\ge d_c$. Here, the value of $d_c$ is 2. (b) Frequencies of the three rowers as a function of $d_{23}$. Bifurcation occurs at $d_{23}=d_c=2$. For very large $d_{23}$, rower 3 becomes almost independent of the other rowers and oscillates with its natural frequency $\omega_0$ while  the first 2 rowers oscillate with the collective frequency of a 2-rowers system $\omega_{coll,2}$. The data here is obtained from a single configuration. Indeed, we checked that the steady state does not depend on the initial configuration if the simulations  does not start with a peculiar condition (such as: all rowers are in phase or all are in anti phase at t=0).}
\label{fig:fig5}
\end{figure}

For a given value of $d_{23}$, we solve the dynamical equations for 3 rowers numerically. We find that, in the dynamical steady state, the first two rowers always oscillate in anti-phase with each other  and the behavior of the third depends on the value of $d_{23}$. In \fref{fig:fig5}(a), we plot the phase difference  $\Delta\phi_{23}=\phi_3-\phi_2$ as a function of time $t$ for various $d_{23}$. We see that, for $d_{23} <d_c=2$, the phase difference $\Delta\phi_{23}$ is constant and for $d_{23} \geq d_c$, $\Delta\phi_{23}$ grows and is modulated in time. The latter implies that the phase locking between the second and third is lost due to the appearance of phase drifting \cite{Adler46,Stratonovich63, Pikovsky03}.

As a result, for $d_{23} < d_{c}$ all 3 rowers oscillate with the same frequency and for $d_{23}\ge d_{c}$, rower 1 and 2 oscillate together with the same frequency but rower 3 oscillates with a different frequency. In \fref{fig:fig5}(b), we plot the frequency of the 3 rowers $\omega_1$, $\omega_2$, and $\omega_3$ as a function $d_{23}$. For $d_{23}=1$, all the frequencies of all  rowers are the same and for $d_{23}\geq 2$, $\omega_1=\omega_2\neq\omega_3$.

Hence, for the set of parameters used in our simulation  ($k=1$, $s=0.8$ and $\alpha=0.1$), the value of the $d_c$ is $2$. We simulated the 3-rowers system for several values of hydrodynamical coupling  $\alpha=0.1, 0.05$, and $0.01$ with the same value of the driving parameters $k$ and $s$ and find $d_c=2$ in all  cases. In this range of investigated parameters,  $d_c$ is independent of $\alpha$. By studying several cases with different sets of driving forces of the rowers for a given $\alpha$, we found that the value of $d_c$ depends on the force constant $k$, and on the amplitude of the oscillation $s$ i.e. $d_c = d_c(k,s)$ (see Appendix~II). This result is very interesting. It means that hydrodynamic interaction strength does not have any role on determining $d_c$. The value of $d_c$ is therefore completely determined by the internal activity of the cilia and not by the fluid viscosity. However, this result is weakened for $N>3$. For $N=4$, we observe $d_c$ depends on $\alpha$ as well as initial configuration of rowers (see Appendix~II).

\begin{figure}[]
\centering
\mbox{
\hspace{-1cm}
\includegraphics[scale = 0.24,angle=270]{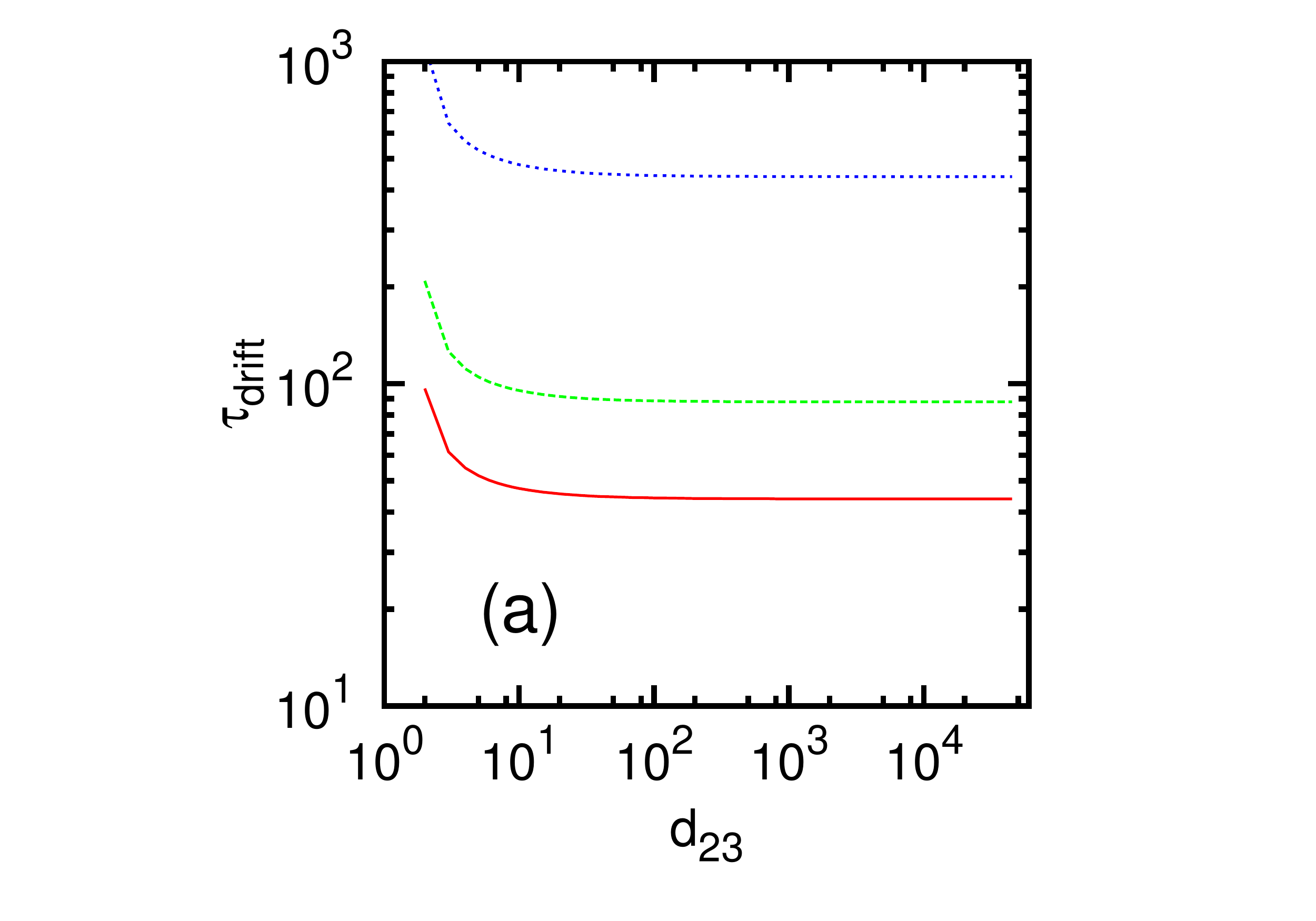}
\hspace{-2cm}
\includegraphics[scale = 0.24,angle=270]{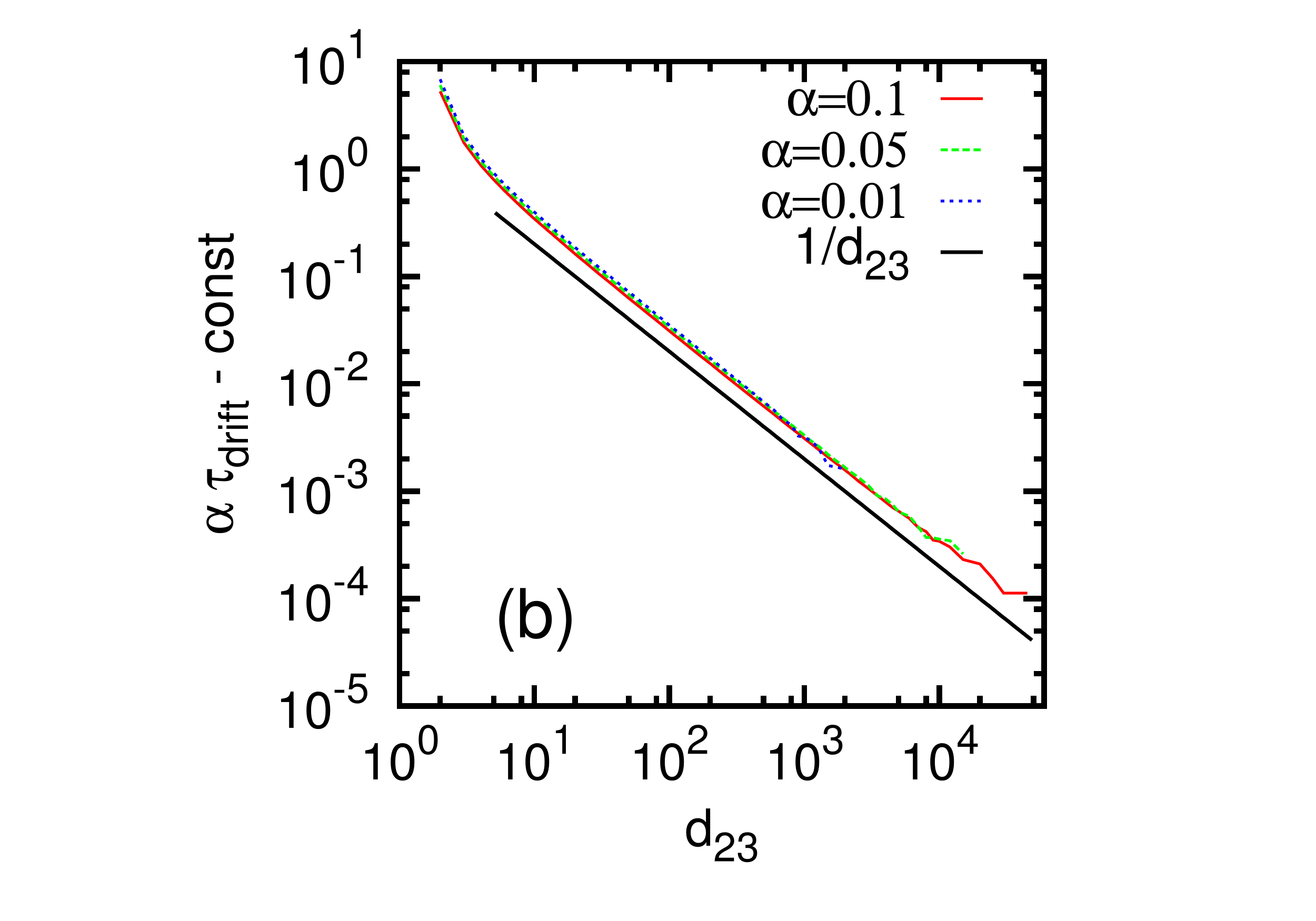}
}
\caption{(a) The log-log plot of $\alpha \tau_{\rm drift}- const$ for $\alpha=0.1, 0.05$ and $0.01$. (b) This scaling makes all the curves fall into a single curve $f(d_{23})$.  The function $f(d_{23})$ s well fitted by $1/d_{23}$. To compute the power laws more accurately, we choose very small step size for the Euler integration method, $h=10^{-5}$.}
\label{fig:fig6}
\end{figure}

The values of  $\omega_1$, $\omega_2$, and $\omega_3$ depends on $d_{23}$. As we have noted above, $\omega_1 = \omega_2$ for all $d_{23}$, but $\omega_3$ behaves differently for $d_{23}\geq d_c$. For very large $d_{23}$, the hydrodynamic coupling of the 3rd rower due to the first two  can be neglected, hence the frequencies of the rowers become independent of $d_{23}$. In this limit, the 3rd rower beats with the natural frequency $\omega_0$ and the first two rowers oscillate with the collective frequency of two rowers $\omega_{coll,2}$ (see section III.1). On the contrary, for small $d_{23}$, the dependence of the rower frequencies on $d_{23}$ is non-trivial. We use the following fits: $\omega_1(d_{23})=\omega_2(d_{23}) \simeq \omega_{coll,2} + \alpha/d_{23}$, and $\omega_3(d_{23}) \simeq \omega_0 -\alpha/d_{23}^2$. In \fref{fig:fig5}(b), we plot these functions (solid for $\omega_3$, and dotted line for $\omega_1$). The two lines match nicely with numerical data (points). Note that $\omega_3$ saturates rapidly to $\omega_0$ ($\omega_3- \omega_0 \simeq \alpha/d_{23}^2$), while the decay of $\omega_1$  to $\omega_{coll,2}$ is slow ($\omega_1-\omega_{coll,2}\simeq \alpha/d_{23}$).  The expression of $\omega_1(d_{23})$ and $\omega_3(d_{23})$ can be understood by the following way. Consider that the 3 rowers oscillate with the same frequency, the  first two being exact anti-phase, and third one being exact anti-phase with the 2nd one. Although this is obviously a very crude assumption, one can use Eq.~\ref{eqn:dynamics} and solve it for the 1st rower:
\bea
\omega_1 \!=\! \omega_0 (\!1-\alpha + \alpha/(d_{23}+1)\!)
         \simeq \omega_{coll,2} + \omega_0 \alpha/d_{23}. 
\eea
Solving the equation for the 3rd rower leads to:
\bea
\!\!\! \omega_3\! =\! \omega_0 (\!1-\alpha/d_{23} + \alpha/(d_{23}+1)\!)
         \simeq \omega_{0} - \omega_0 \alpha/d_{23}^2. 
\eea

The growth of $\Delta\phi_{23}$ is not uniform and  follows a periodic pattern  in time (\fref{fig:fig5}(a)). For $d_{23}=2$, the phase difference rather shows sharp $2\,\pi$ jumps followed by almost flat regimes. With the increase of $d_{23}$, the growth rate of the phase difference becomes faster and steps disappear. In order to characterize the nature of the phase drift, we define the time scale $\tau_{\rm drift}(\alpha, d_{23})$ as the time needed for the phase difference $\Delta\phi_{23}$ to increase by $2\,\pi$ (i.e. the time period of $\Delta\phi_{23}$), and we compute it for a given internal activity of rowers (i.e for fixed $k$ and $s$). We find that $\alpha\,\tau_{\rm drift}(\alpha, d_{23}) = f(d_{23}) + const$, where the function $f$ does not depend on $\alpha$. In \fref{fig:fig6}, we plot   $\alpha \tau_{\rm drift} - const$ in log-log scale and we show that the curves for different $\alpha$ collapse into a single curve. The scaling function  $f(d_{23})$ decays algebraically to zero as $1/d_{23}$. This collapse and the $1/d_{23}$ decay can be rationalized using the following simple arguments and approximations. From the definition of the drift time, it is easy to see that $\tau_{\rm drift} (d_{23}) \sim 1/(\omega_3 (d_{23}) -\omega_1 (d_{23})) \simeq 1/(\alpha\, d_{23}) + const/\alpha$, using the fitting expressions for $\omega_1(d_{23})$ and $\omega_3(d_{23})$.

\subsection{3. 4-rowers case}

The  phenomena of phase drifting and bifurcation is also observed for a system of 4 or more rowers. Here, we study the case of 4 rowers. We divide the 4 rowers into two subgroups in order to study  different configurations. For a given internal structure (i.e. for given $k$ and $s$), one obtains a variety of behavior, as the value of $d_c$ depends on the number of rowers  in each group

\begin{figure}[h]
\centering
\mbox{
\hspace{-1.5cm}
\includegraphics[scale = 0.18,angle=0]{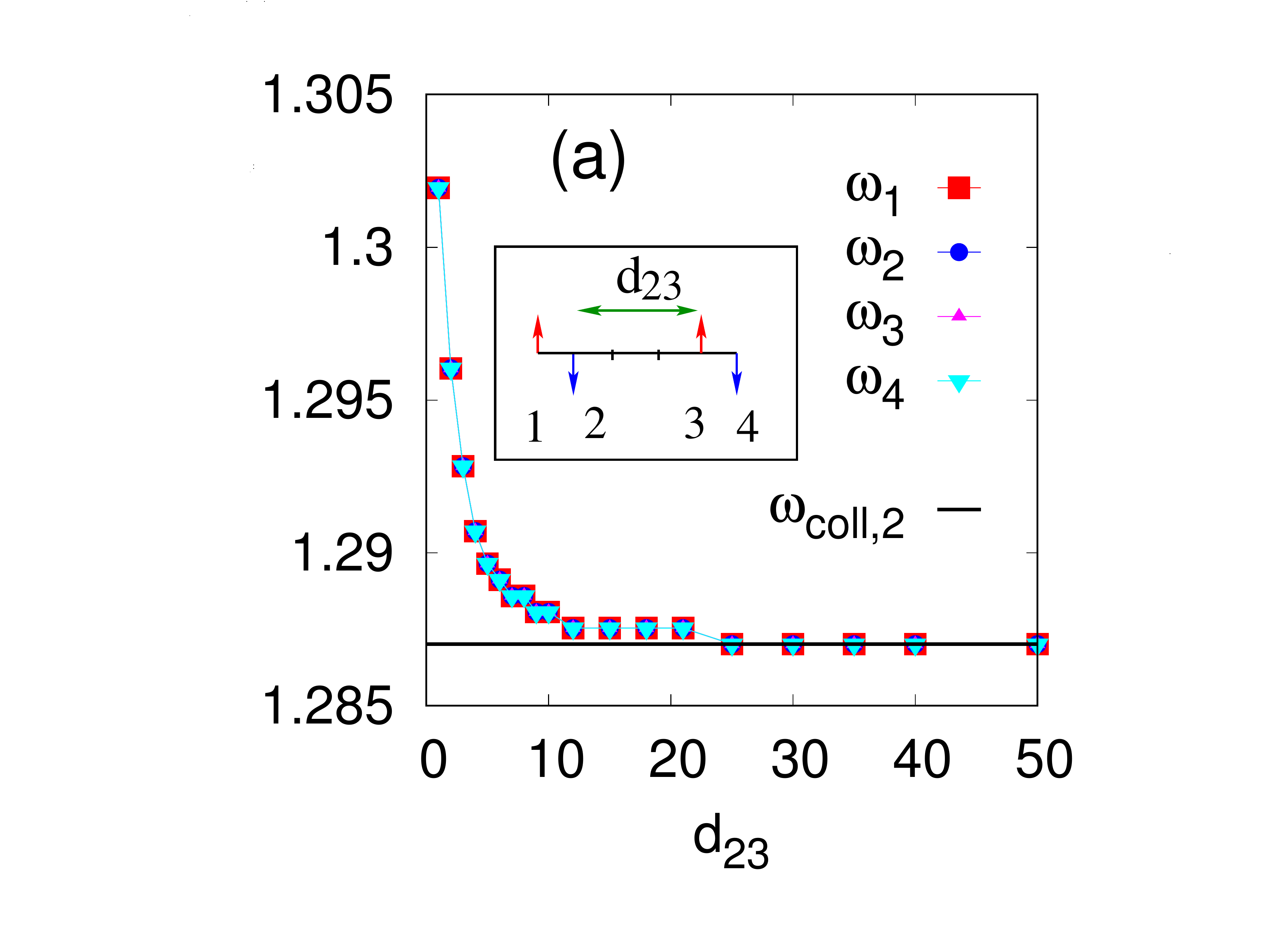}
\hspace{-2.5cm}
\includegraphics[scale = 0.25,angle=0]{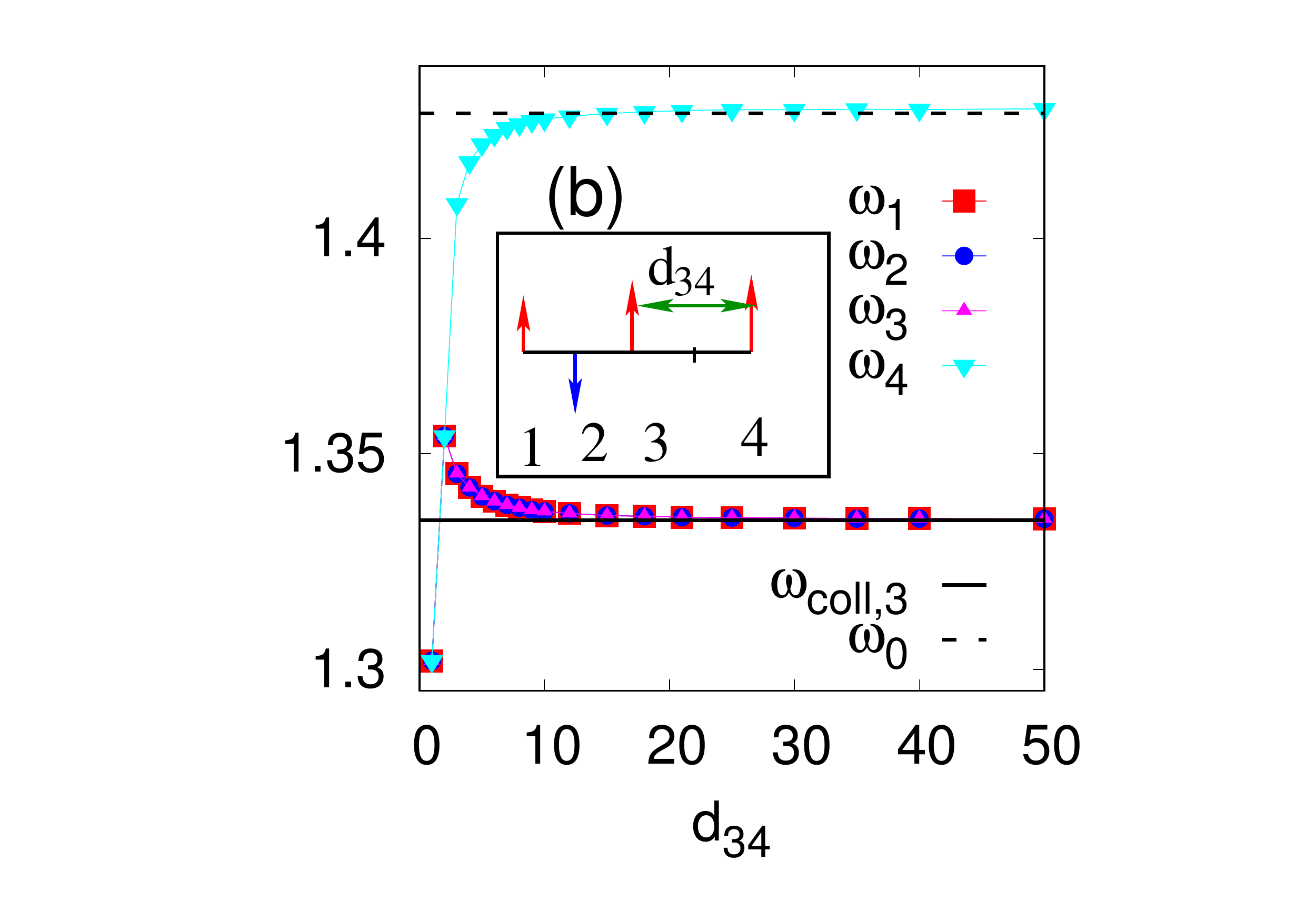}
}
\caption{Angular frequencies of 4 rowers for different gaps between clusters;  $\alpha=0.1$ : (a) symmetric case 2-2, no bifurcation is observed, (b) asymmetric case 3-1, a bifurcation observed.}
\label{fig:fig7}
\end{figure}

Case I --- Asymmetric case: The first group of rowers has 3 consecutive rowers with  equal gap, $d_{12}=d_{23}=1$. The second group consists only of the 4th rower and the distance separating the two subgroups is $d_{34}$. In \fref{fig:fig7} (b), we plot the frequencies for rowers as a function of $d_{34}$. A bifurcation similar to the one obtained for three rowers is observed for $d_{34} \geq d_c$. The critical distance for this case is different from the 3-rowers system. Unlike for $N=3$, the value of $d_c$ depends $\alpha$ and initial configuration of rowers (see Appendix~II). For very large $d_{34}$, the 3-rowers group behaves independently of the 4th rower which oscillates at $\omega_0$, the natural single rower frequency.

Case II --- Symmetric case: The first group has 2 consecutive rowers with  gap, $d_{12}=1$. The second group consists of the
3rd and 4th rower with gap $d_{34}$=1. Here, all rowers oscillate with the same frequency and as the distance 
between two groups $d_{23}$ is very large the frequency asymptotically saturates to the frequency of 2-rowers
system. Here the two clusters independently beat with the same frequency characteristic of their size (\fref{fig:fig7} (a)).

\begin{figure}[]
\centering
\mbox{
\hspace{-1cm}
\includegraphics[scale = 0.25,angle=270]{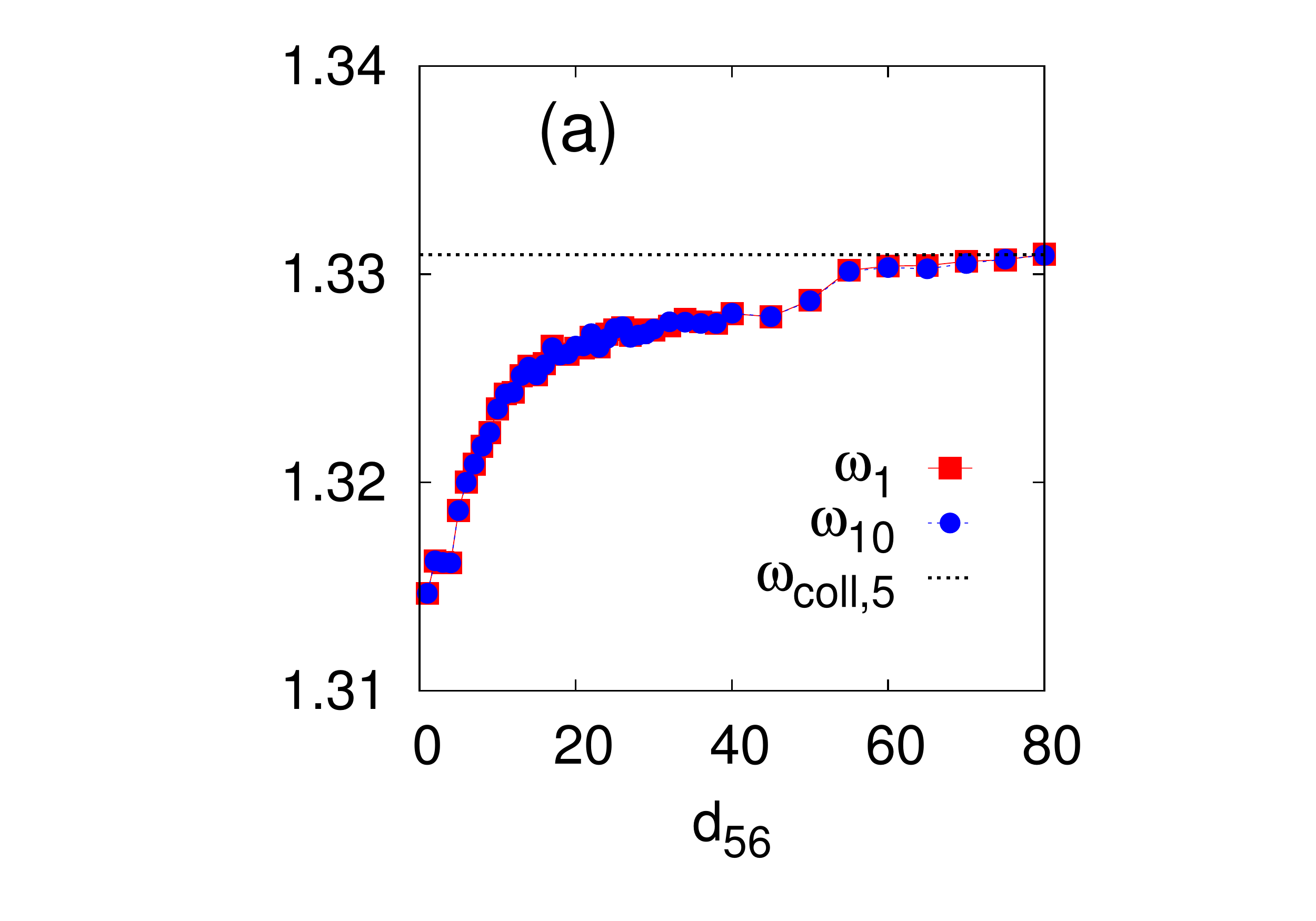}
\hspace{-2cm}
\includegraphics[scale = 0.25,angle=270]{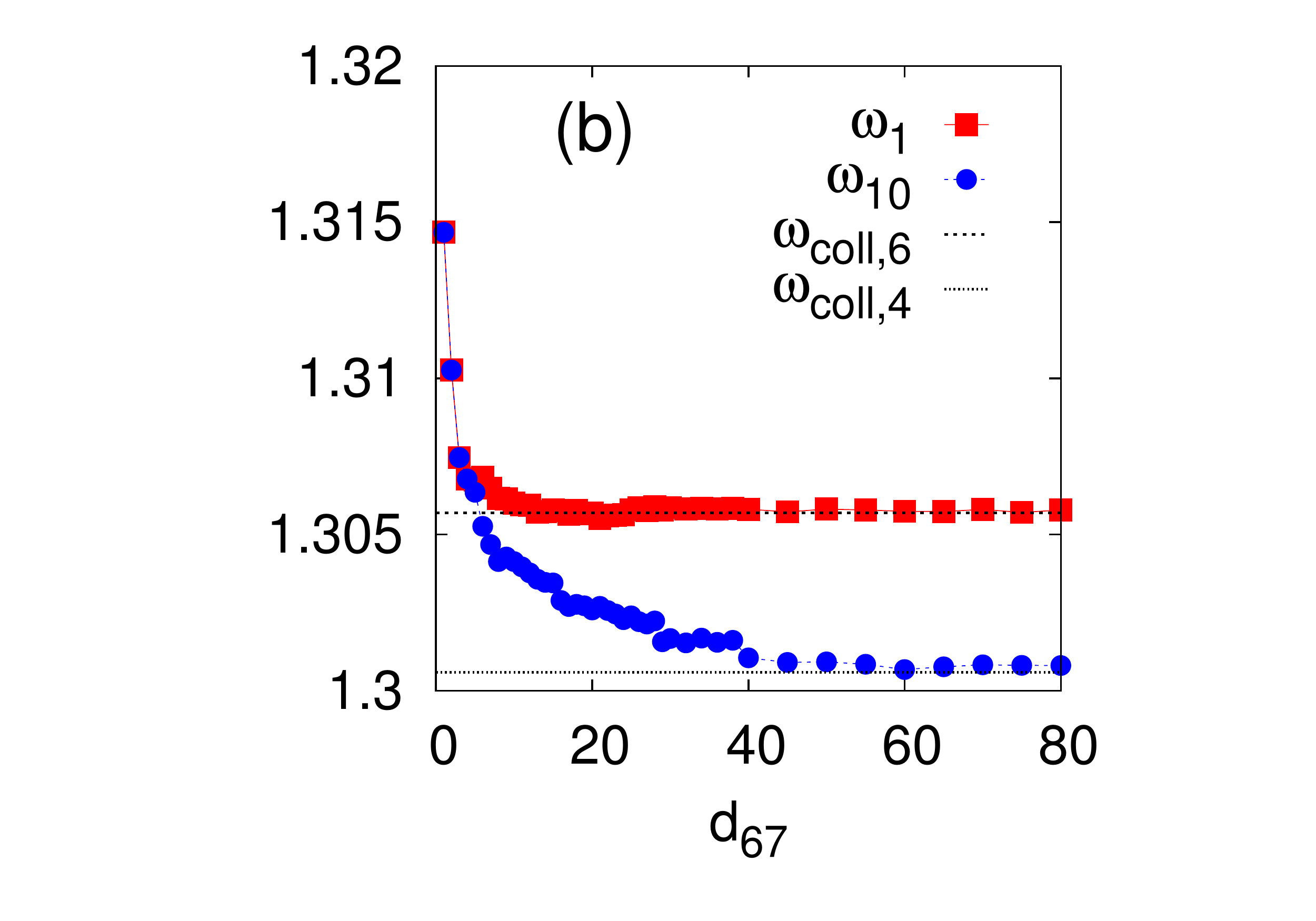}
}
\mbox{
\hspace{-1cm}
\includegraphics[scale = 0.25,angle=270]{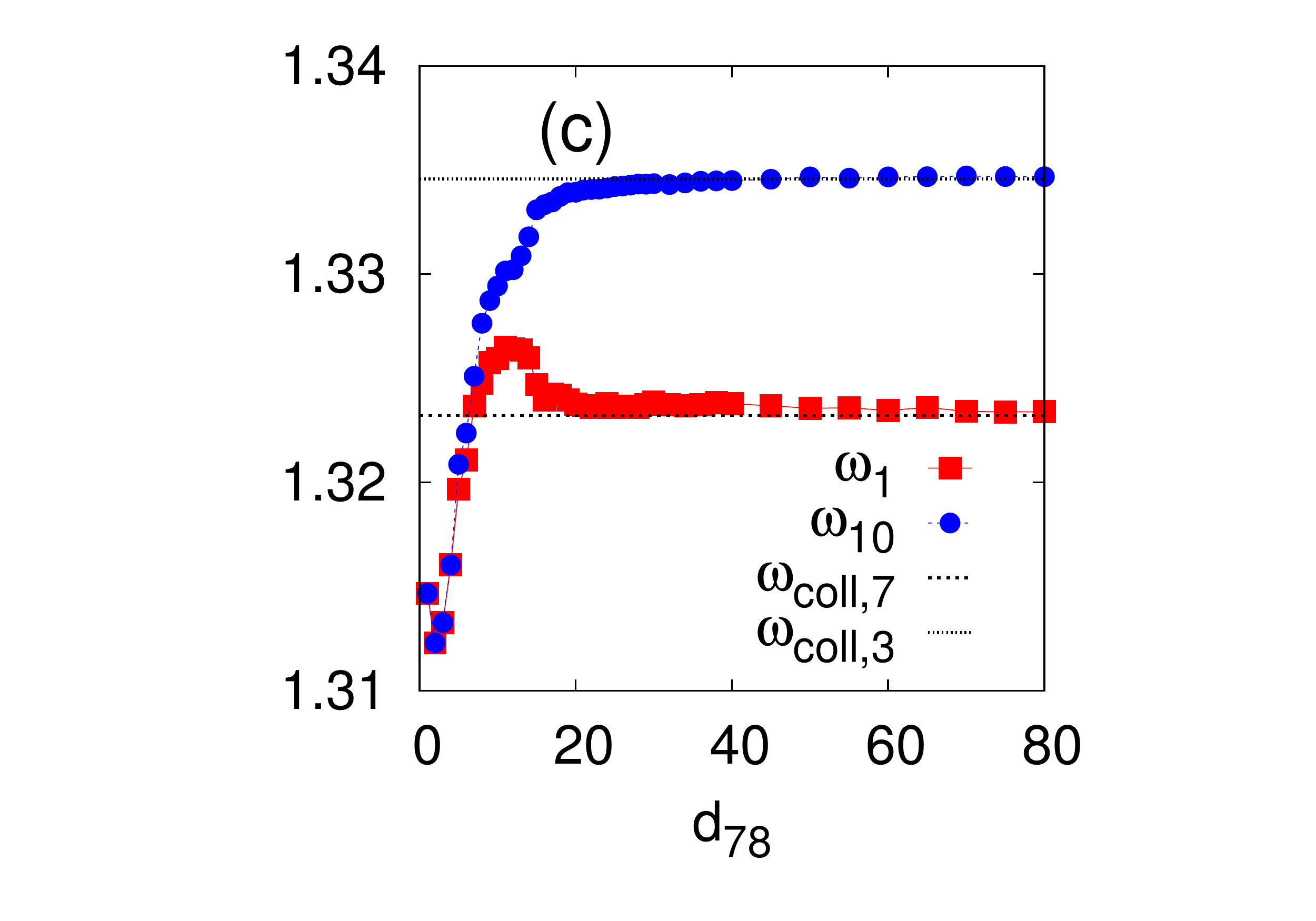}
\hspace{-2cm}
\includegraphics[scale = 0.25,angle=270]{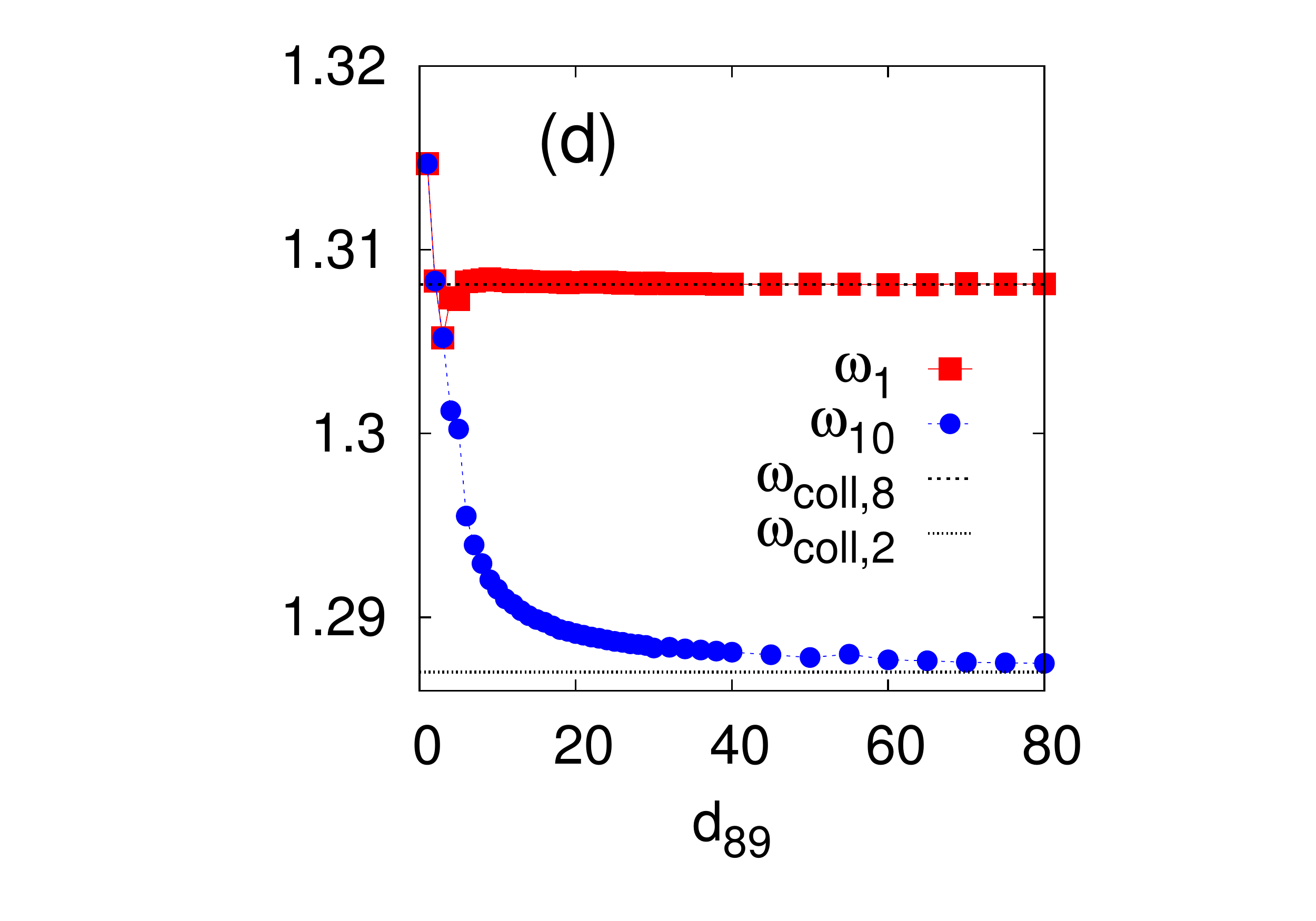}
}
\includegraphics[scale = 0.25,angle=270]{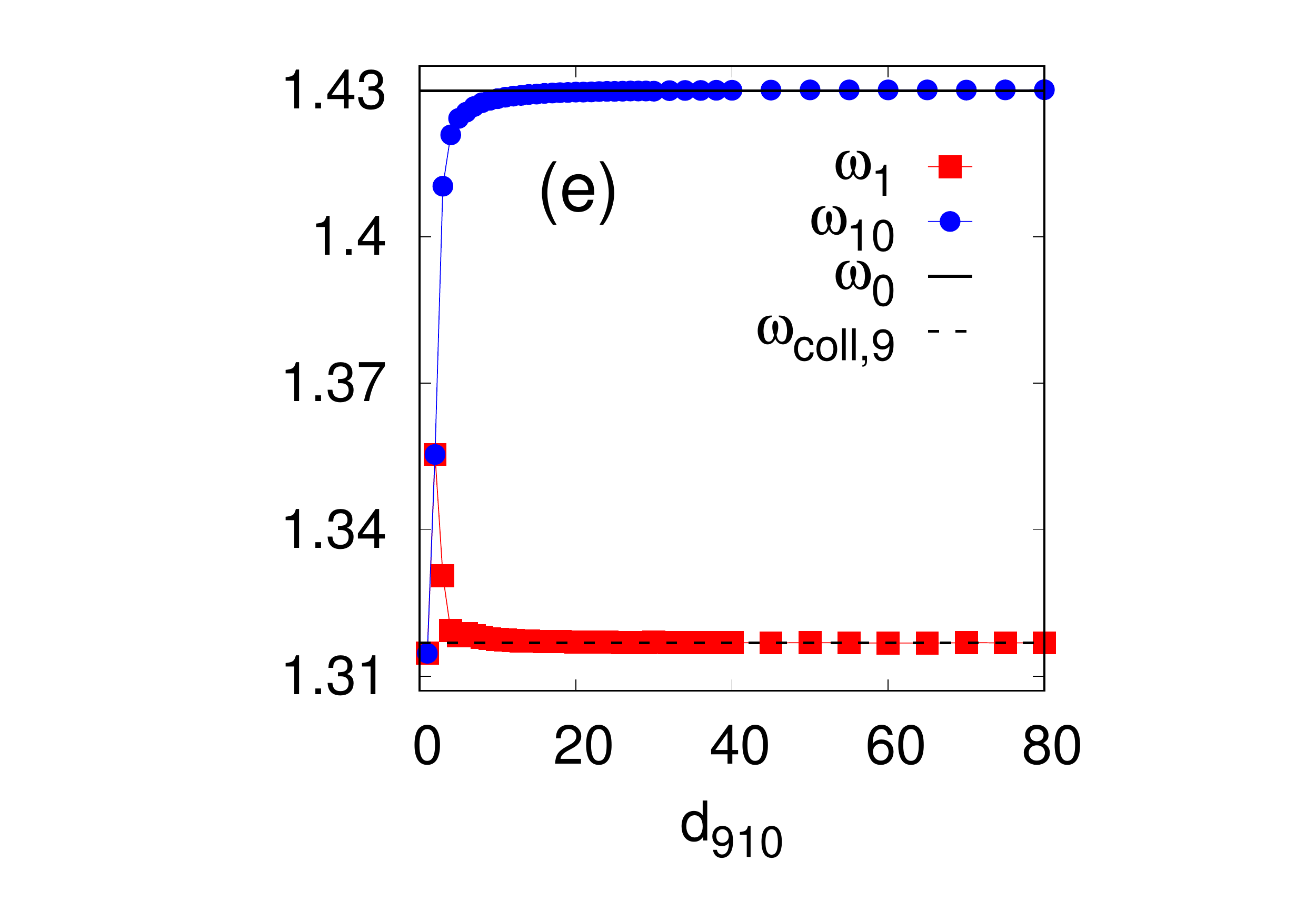}
\caption{ From (a) to (e), the sizes of the 2 clusters are respectively 5-5/ 6-4/ 7-3/ 8-2/ 9-1. When the size of the two clusters is comparable (5-5 and 6-4) the solution of the dynamical equations depends on initial conditions. The data are then averaged over $1000$ initial conditions.}
\label{fig:fig8}
\end{figure}

\subsection{4. 10-rowers case}

We divide the 10 rowers set into two subgroups  and studied the 5 different following cases.  

Symmetric case  (5-5):  each subgroup consists of 5 rowers equally spaced, but the distance $d_{56}$ between the two groups can vary.
 We observe no bifurcation 
in this symmetric case (see \fref{fig:fig8} (a)). 

There are 4 asymmetric cases: 6-4, 7-3, 8-2 and 9-1 groups. In all these cases,
a bifurcation occurs. In \fref{fig:fig8} (b), (c), (d), (e), results for the collective frequencies of the 2 clusters are shown. The critical distances after which bifurcation occurs are different in each case.

From the above study of finite size systems, we learn  two instructive features. (i) As expected from symmetry, there is no bifurcation if the sizes of the 2 clusters are equal, instead all rowers beat at the same frequency whatever the distance between the 2 clusters. (ii) The distance where the bifurcation occurs is maximal for non-trivial cluster sizes and is minimal for strikingly different sizes.

\begin{figure}[]
\centering
\mbox{
\hspace{-1cm}
\includegraphics[scale = 0.25,angle=270]{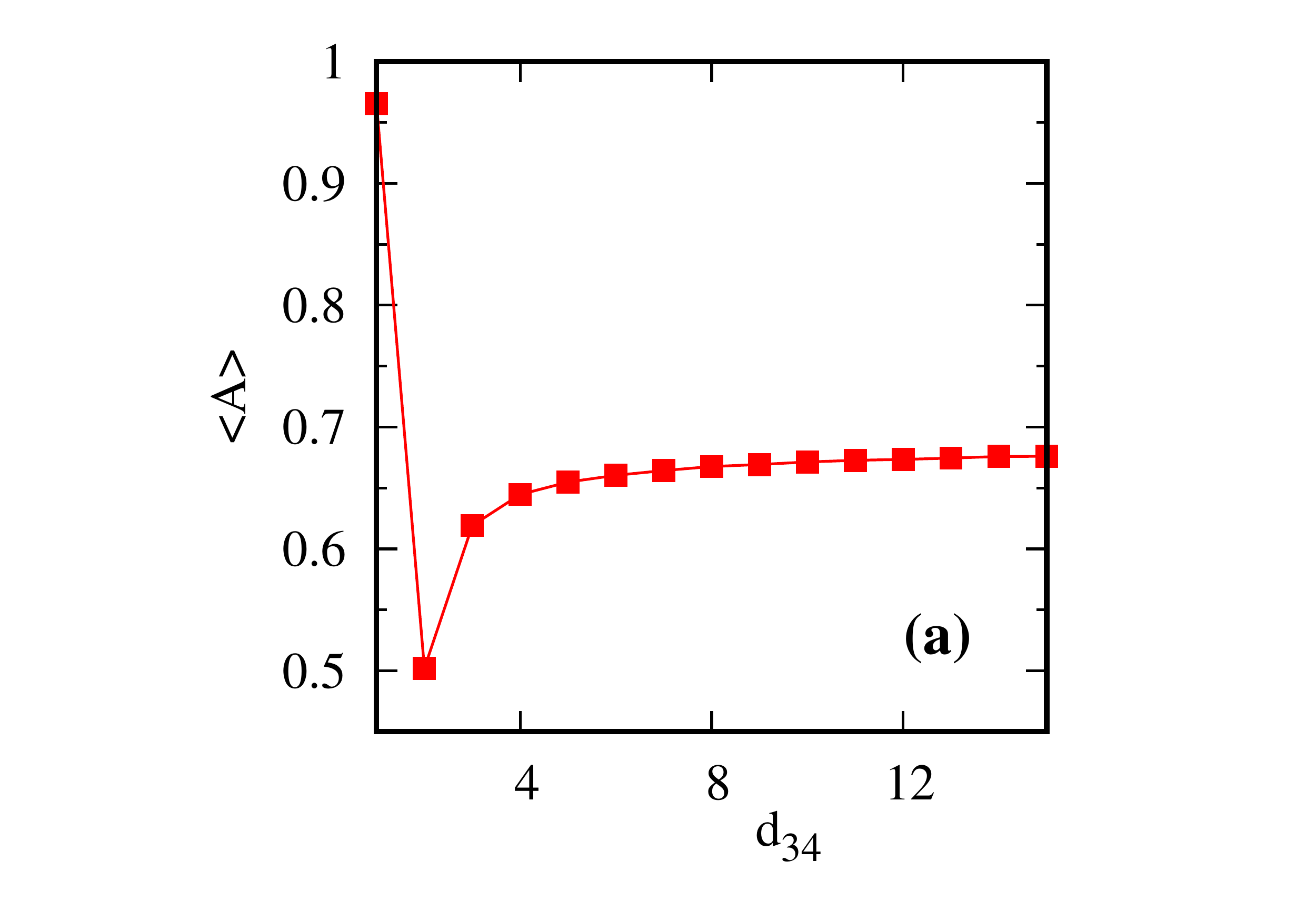}
\hspace{-2cm}
\includegraphics[scale = 0.24,angle=270]{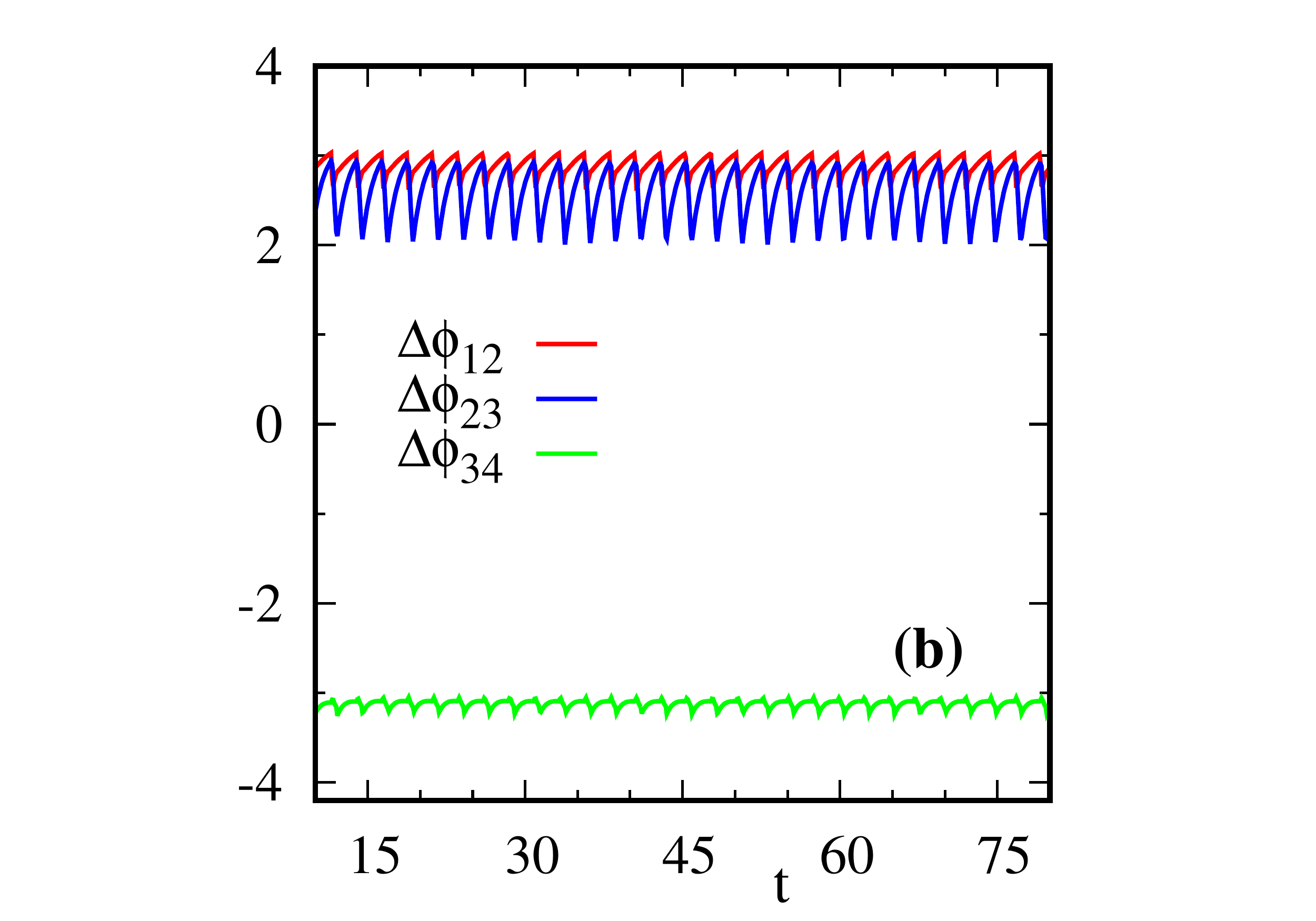}
}
\mbox{
\hspace{-1cm}
\includegraphics[scale = 0.24,angle=270]{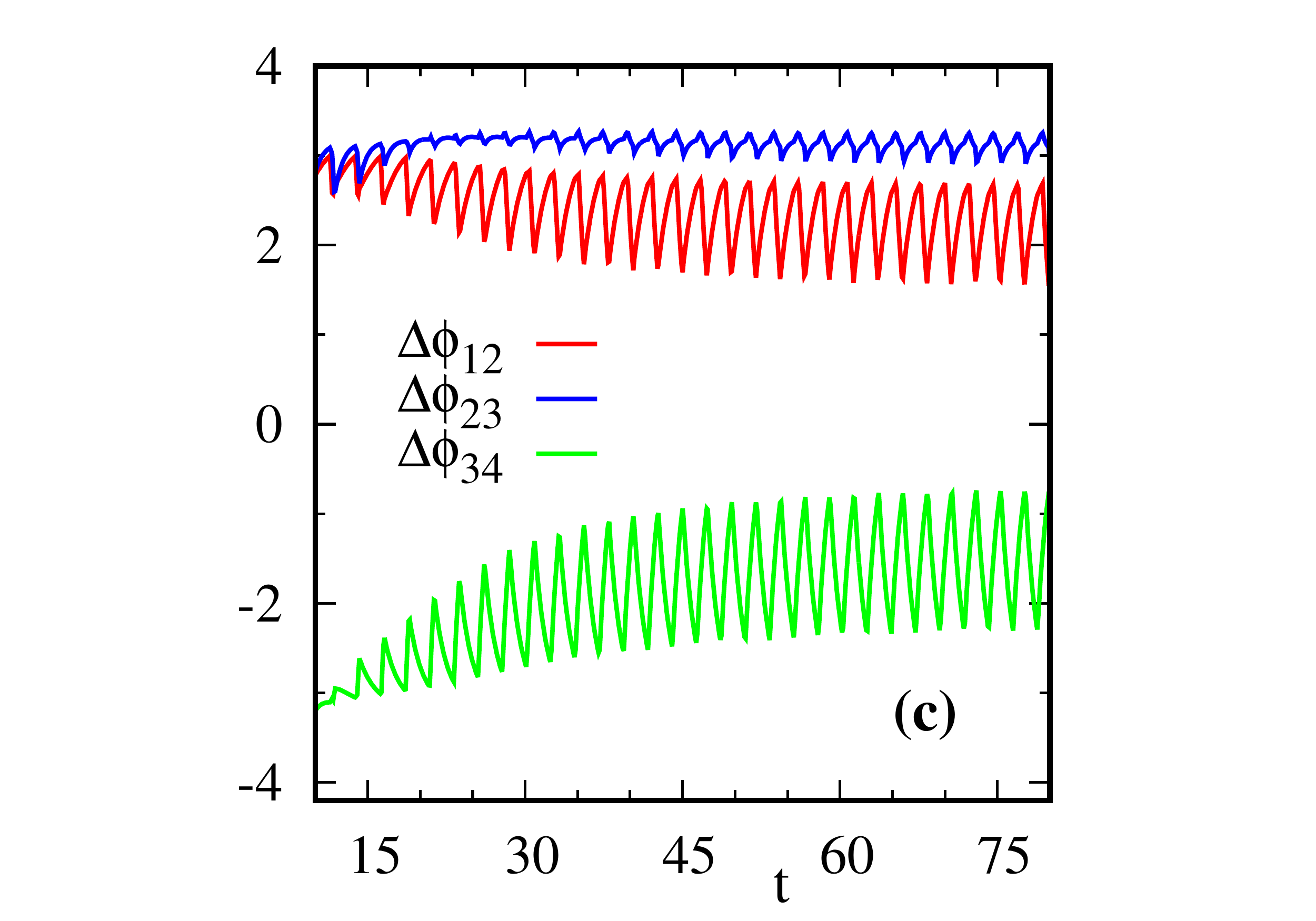}
\hspace{-2cm}
\includegraphics[scale = 0.24,angle=270]{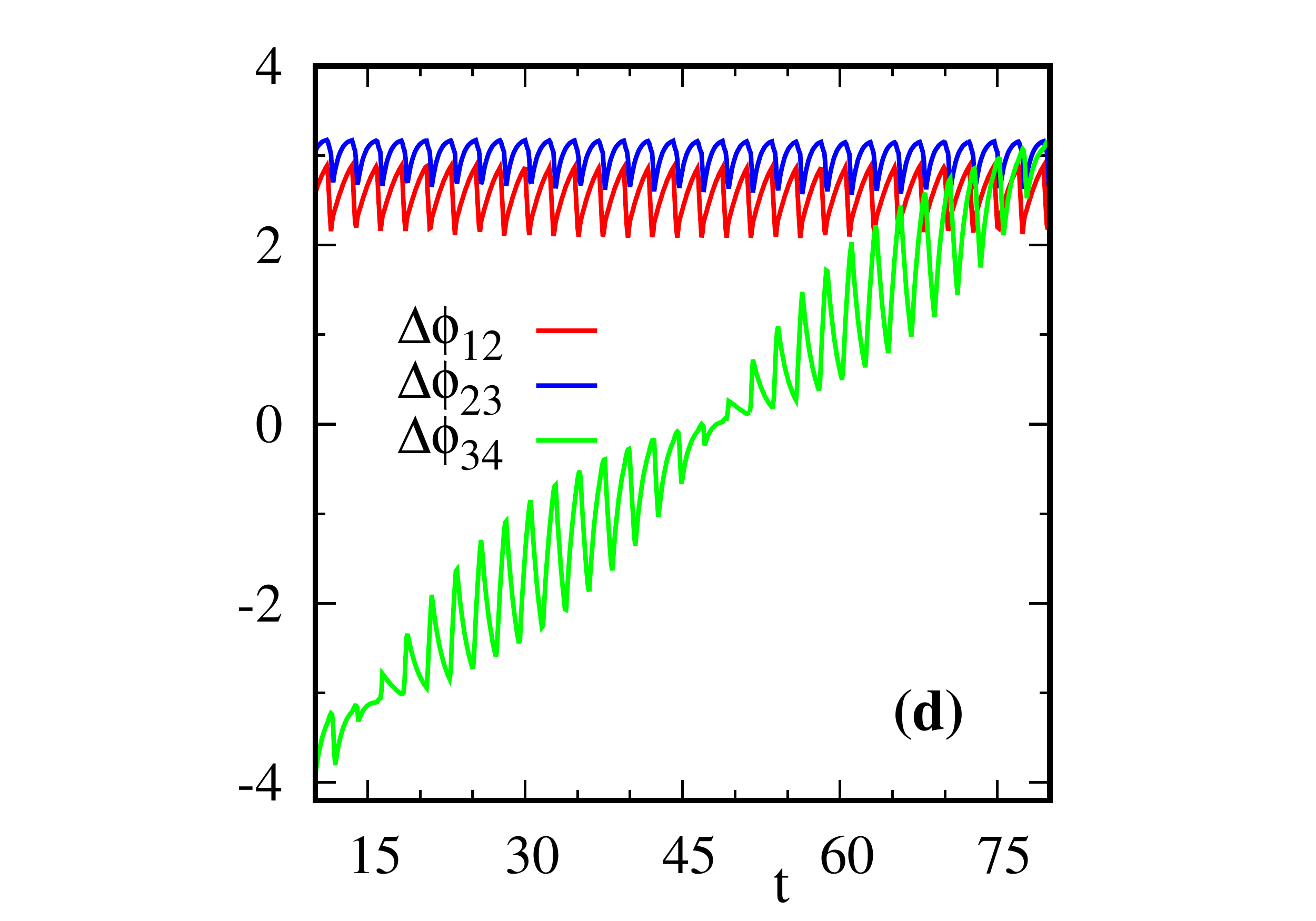}
}
\caption{
(a) The time averaged order parameter $\langle A\rangle$, averaged over a large time window after the system reaches dynamical steady state for an arbitrary initial condition for the 3+1-rowers case, is plotted against separation $d_{34}$. The time evolution of phase differences between consecutive rowers for $d_{34}=1$, $d_{34}=2$, and $d_{34}=50$ are plotted in (b), (c), and (d) respectively.}
\label{fig:fig9}
\end{figure}

\subsection{5. Link between the order parameter $A$ and the rowers phases $\phi$ }

As seen before, local observables such as phase differences between rowers $\Delta \phi_{ij}$ or individual frequencies $\omega_i$ are necessary to describe in detail the good or poor coherence state of an assembly of rowers. If one now looks at the global observable $A$, how will $A$ be related to the  dynamical state of the system? The relation between $\Delta \phi_{ij}$ between consecutive rowers and  $A$  is given by \eref{eqn:orderpara}.  However,  by looking only at the value of $A$,  it is not possible to infer the dynamical state of the system (i.e $\Delta \phi_{ij}$).  The study of the 3+1-rowers case is instructive in this respect and is reported here. As seen in \fref{fig:fig7} (b), $d_c=3$ in this case. We show in \fref{fig:fig9} that 3 distinct dynamical phases can be identified by comparing the time averaged value  of $A$ with the rowers phase differences. If $d_{34}=1$, the group of rowers is compact and oscillate in  almost anti phase, hence  $\langle A\rangle$ is maximal and close to $1$. If $d_{34}=2$, the system stands right before the bifurcation ($d_c=3$), and the rowers state is dynamically disordered as one can see on \fref{fig:fig9} (c). If $d_{34}$ is much larger than $d_c$ (and this is illustrated by looking at $d_{34}=50$, the two clusters of rowers oscillate independently but the 3 rowers in the first group are coordinated in almost anti phase, leading to $A$ reaching a saturation value after the dip at $d_{34}=2$.

\subsection{6. Collective beating  frequency of rowers on  a regular array }

For a regular array of $N$  equally spaced rowers, we observe that all rowers oscillate at same frequency (except a few rowers close to the boundaries for very large $N$, where $N$ is the number of total rowers) showing  phase coherence both for $\alpha<0$ (in-phase synchronization) , or  for $\alpha>0$ (anti-phase metachronal waves) \cite{CosentinoPre03}. We studied here how the collective frequency $\omega_{coll}$ depends on $N$. We observe that for $\alpha<0$, $\omega_{coll}$ strongly depends on $N$ (see Appendix IV). For $\alpha>0$, however $\omega_{coll}$  stabilizes to a constant value. This confirms that metachronal waves in mucociliary systems can be stable even in large samples.

\begin{figure}[h]
\centering
\includegraphics[scale = 0.3,angle=270]{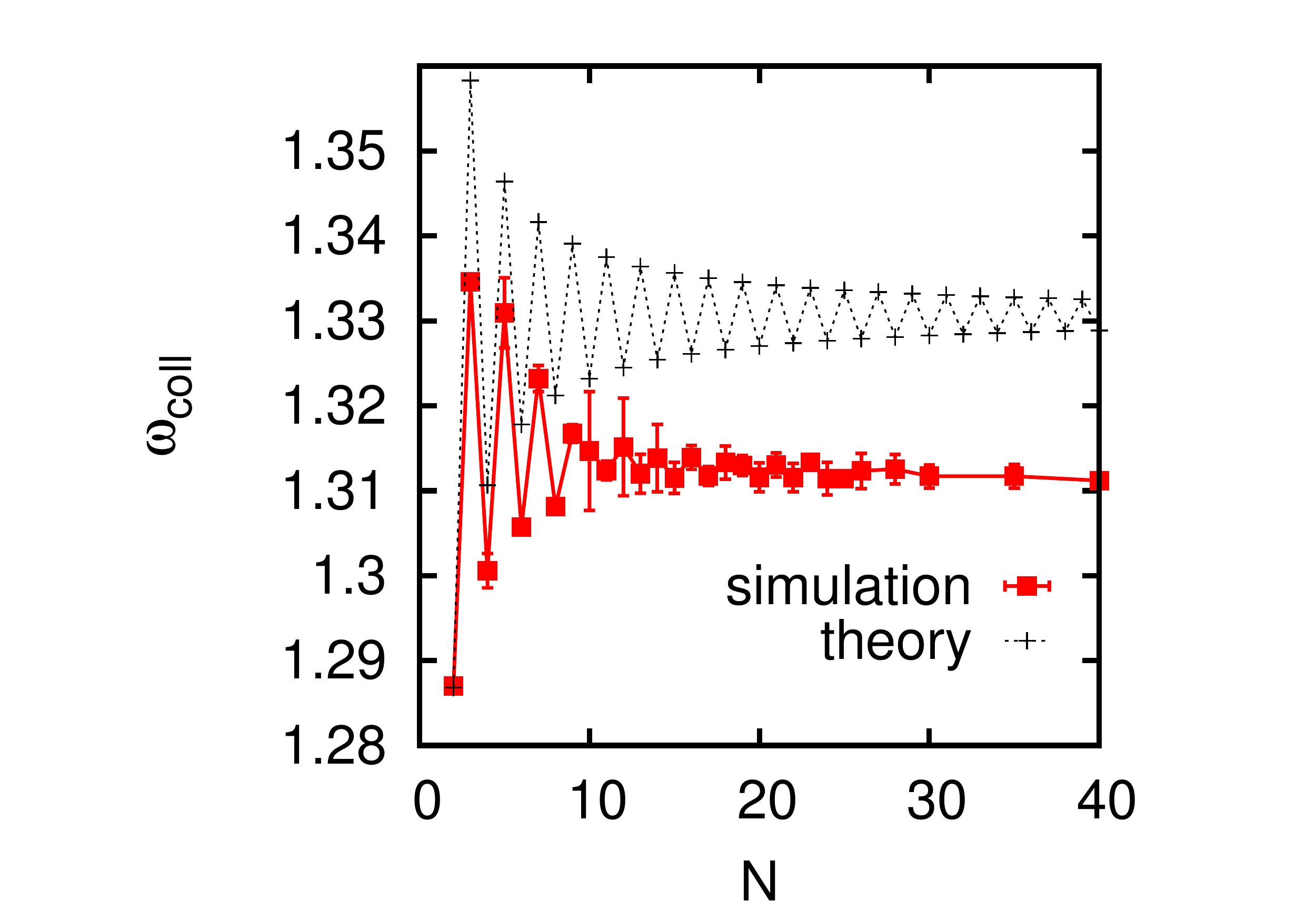}
\caption{The collective frequency is plotted as a function of total number of rowers $N$. The rowers are placed on a regular lattice with lattice constant $d=1$. The data are averaged over $1000$ initial configurations. It reaches a stable frequency for $N>N_c$ ($\sim 15$). This can be interpreted as the minimal cluster size for the existence of a metachronal wave at the frequency typical of large systems.}
\label{fig:fig10}
\end{figure}

The value of the collective beating frequency $\omega_{coll}$ of the rowers oscillates for small $N$ (total number of rowers), and it eventually converges to a constant value for large $N$. We can understand this behavior from a very simple picture. Consider the following dynamical steady state for our oscillating rowers in which a rower beats in perfect anti-phase with its nearest neighbours. This implies $y_1(t)=-y_2(t)=y_3(t)=-y_4(t)=.., =y_{N-1}(t)=-y_{N}(t)$. In this particular case, we can easily solve the dynamics (\eref{eqn:dynamics}) and calculate $\omega_{coll}$. For $N$ numbers of total rowers, the collective frequency is given by,
\bea
\!\!\!\omega_{coll,N}\!=\!\omega_0 [1 -\alpha(\!1-\frac{1}{2}+\frac{1}{3}+.. + (-1)^N\frac{1}{N-1}\!)].
\label{eqn:omega_approx}
\eea
The coefficient of $\alpha$ is the alternate harmonic (convergent) series. In \fref{fig:fig10}, we plot  both the collective frequency computed from theory (\eref{eqn:omega_approx}), and the one obtained from simulation. For $N=2$, the values of collective frequency computed from these two different methods match exactly; the assumption of perfect anti-phase synchronization is indeed true for $N=2$ (seen previously, \fref{fig:fig3}(a)). However, for $N>2$, this is not exactly the case because the assumption of perfect anti-phase synchronization is no more valid.  We believe that the influence of boundaries is indeed important even for large $N$, as seen on \fref{fig:fig10}.


We have simulated systems with small number of rowers for another value of $\alpha(=0.2$) and have observed that the qualitative behavior of our results is independent of coupling strength.

\section{IV. Collective behavior of many spatially heterogeneous rowers}


In the previous section, we have studied how spatial heterogeneity of rowers  affect the coherent beating when the number of rowers is small. We observed that the spatial arrangement of rowers can lead to phase drifting, phase incoherence, and bifurcation in frequency when the separation between two clusters is greater than a critical distance. We now turn to asking the question of  how spatial heterogeneities in the position of rowers affect the coherent beating in a system of hundreds of rowers. 

On the one dimensional lattice, we will consider three types of heterogeneities  --- (i)  regular clustered configurations, (ii) random configurations, and (iii)  random clustered configurations, which is an intermediate situation between (i) and (ii).

We  generate these types of heterogeneities for different values of the density of rowers $\rho$, and study the dynamical properties of the corresponding systems.

\begin{figure}[]
\centering
\includegraphics[scale=0.26]{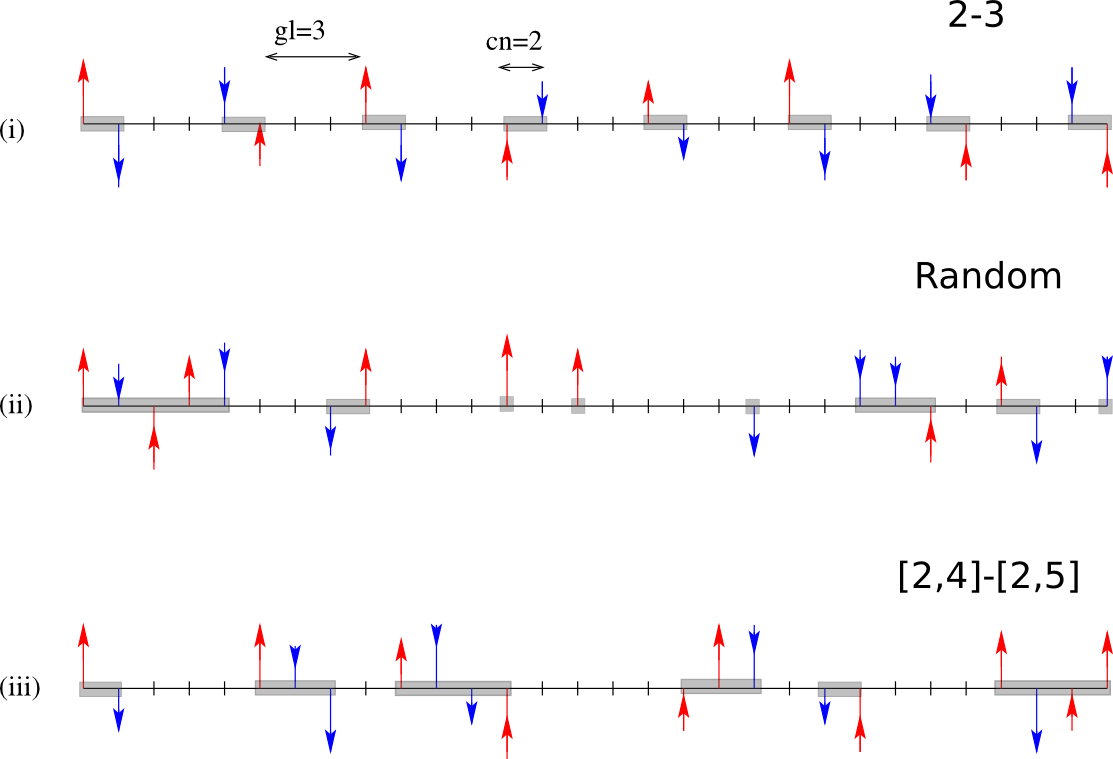}
\caption{ Different types of spatially heterogeneous rowers configurations (rowers are the arrows). (i) Regular clustered: clusters of fixed number of rowers are placed on the lattice  regularly spaced. In this diagram, the number of rowers in a cluster is $cn=2$, and the length of the gap between two clusters of rowers $gl=3$. We refer to this lattice as ``2-3''. (ii) Random: rowers are randomly placed on the lattice. (iii) Random clustered: the number of rowers in a cluster is not fixed, and is chosen uniformly in the interval [$cn_{min},cn_{max}$]. The gap lengths between two consecutive clusters are also randomly taken from [$gl_{min},gl_{max}$]. In this diagram, the minimum and maximum length of clusters of rowers are $cn_{min}=2$ and $cn_{max}=4$. The minimum and maximum length of a gap between two clusters are $gl_{min}=2$ and $gl_{max}=5$. This structure is referred to as ``[2,4]-[2,5]''.
}
\label{fig:fig11}
\end{figure}

(i)  In the case of regular clustered heterogeneity, clusters of fixed number of rowers are placed on a lattice, separated by  regular gaps. A typical lattice configuration of rowers is shown in \fref{fig:fig11}(i). We call $cn$  the number of rowers in a cluster, and $gl$  the gap length between two consecutive clusters. 
We refer to such clusters as ``cn-gl''. One way to generate configurations of different densities for a given cluster length $cn$ is to vary the gap length $gl$
between two clusters. The density  is  given by,
\bea
\rho = \frac{cn}{cn+gl-1}.
\label{eqn:den_struc1}
\eea

(ii)  For random heterogeneity (see \fref{fig:fig11}(ii)), $N$ rowers are placed randomly  on  a one dimensional lattice of size $L$. The density of rowers is given by $\rho=N/L$. For a fixed $N$, we generate configurations with different values of $\rho$.

(iii) In the case of random clustered heterogeneity, we introduce randomness in the sizes of the clusters  as well as in the gap lengths between two clusters. Here, cluster lengths $cn$ are chosen randomly between $cn_{min}$ and $cn_{max}$ using uniform random distribution. The gap lengths $gl$ are chosen randomly between $gl_{min}$ and $gl_{max}$ using uniform random distribution. We refer to  this type of heterogeneity as [$cn_{min},cn_{max}]-[gl_{min},gl_{max}$]. A typical lattice configuration is shown in \fref{fig:fig11}(iii). In order to generate configurations with different densities keeping fixed $cn_{min}$, $cn_{max}$, and $gl_{min}$, we vary $gl_{max}$. The average density can be computed numerically, averaged over a large number of realizations:
\bea
\rho =\langle  \frac{cn} { cn  +  gl  -1}\rangle. \nonumber
\label{eqn:den_struc2}
\eea

In our computation, we will use 2-$gl$ and 3-$gl$ regular clustered heterogeneities. For random clustered configurations, we will use the following heterogeneities: [2,4]-[2,$gl_{max}$] and [3,5]-[2,$gl_{max}$].

We measure the distribution of beating frequency  $P(\omega)$ as a function of $\rho$ in the different types of heterogeneities discussed above. We use $N=100$ and $\alpha=0.1$. In order to compute $P(\omega)$, we consider several spatial structures and initial conditions of rowers. In the case of random and regular clustered, we take $100$ different spatial configurations of rowers. For regular clustered heterogeneity, as the spatial configuration of rowers on a lattice is fixed, we consider $100$ different initial configurations of the rowers.

\begin{figure}[]
\centering
\mbox{
\includegraphics[scale = 0.3,angle=270]{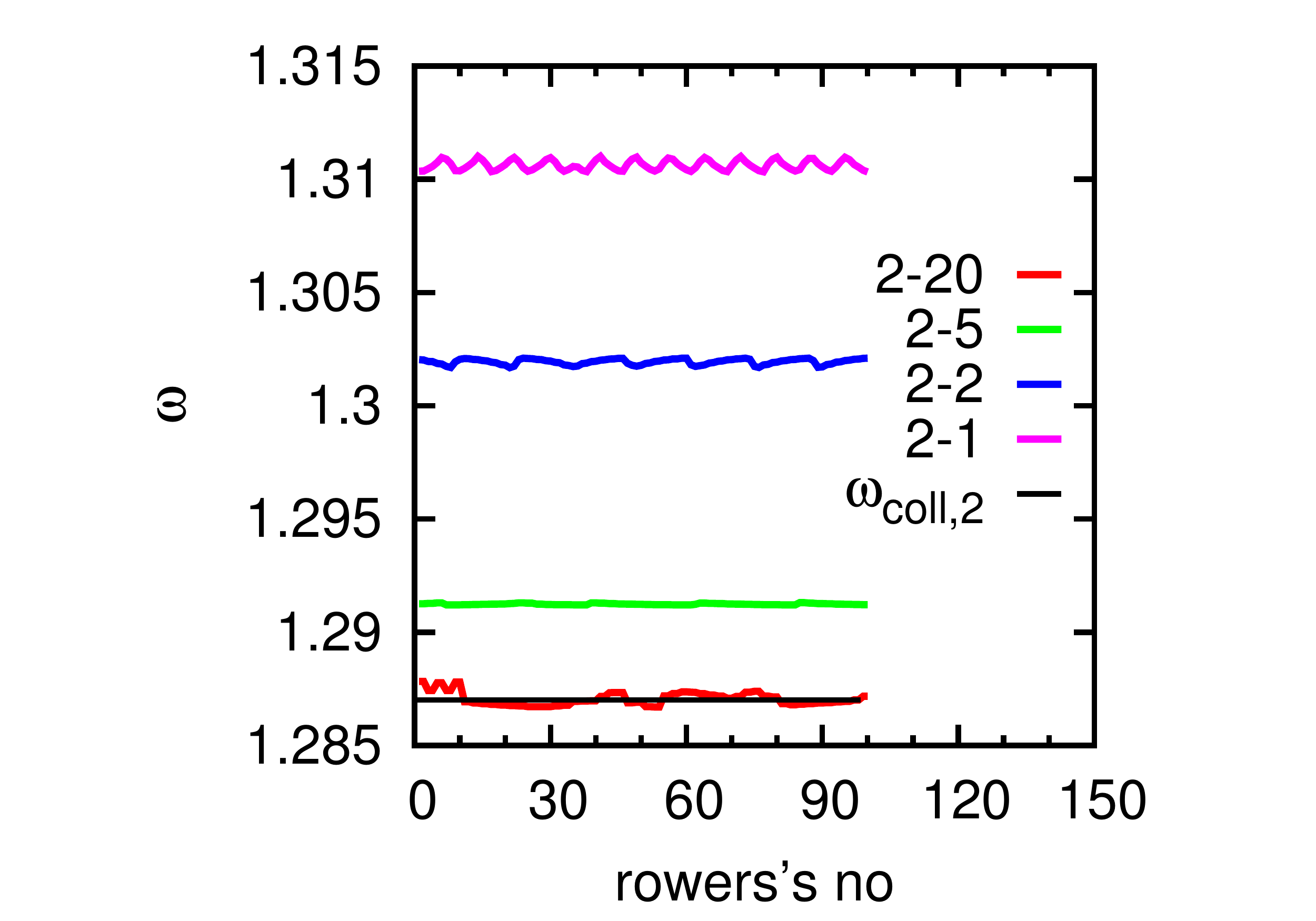}
}
\caption{(i)  Regular clustered case. Spatial profiles of the rowers frequency  for ``2-gl'' regular clustered heterogeneity for different gap lengths $gl=1,2,5$ and $20$. The densities corresponding to these $gl$ values are given by $\rho=$ 1, 0.67, 0.33, and 0.1 respectively. For a given gap length, all rowers in a lattice oscillate approximately  in a single collective frequency. The value of the collective frequencies depend on the value of $gl$. For $gl=20$, the inter-clusters hydrodynamic coupling can be neglected, and in this limit, rowers oscillate with the collective frequency of a cluster of two  rowers, $\omega_{coll,2}$.}
\label{fig:fig12}
\end{figure}

\subsection{1. Case $(i)$: Regular clustered }

We consider ``2-gl'' regular clustered configurations. In \fref{fig:fig12}, we plot the spatial profile of the average beating frequency. We observe that for a given $\rho$ all rowers oscillate with a single collective frequency. The value of this collective frequency decreases as  $\rho$ is decreased (by increasing $gl$). In the large $gl$ limit, the inter clusters interaction can be neglected. As a consequence, we see in \fref{fig:fig12} that the frequency of the rowers for $gl=20$ is the same as the collective frequency of a cluster of two rowers separated by a distance $d=1$, $\omega_{coll,2}$ (section IV.1).  However, for ``3-gl'' regular clustered configurations, $\omega$ increases as $\rho$ is decreased, and reaches $\omega_{coll,3}$ at small $\rho$ (see Fig.~\ref{fig:fig15}(a)).

\subsection{2. Case $(ii)$: Random }

\begin{figure}[htb]
\centering
\mbox{
\hspace{-1cm}
\includegraphics[scale = 0.23,angle=270]{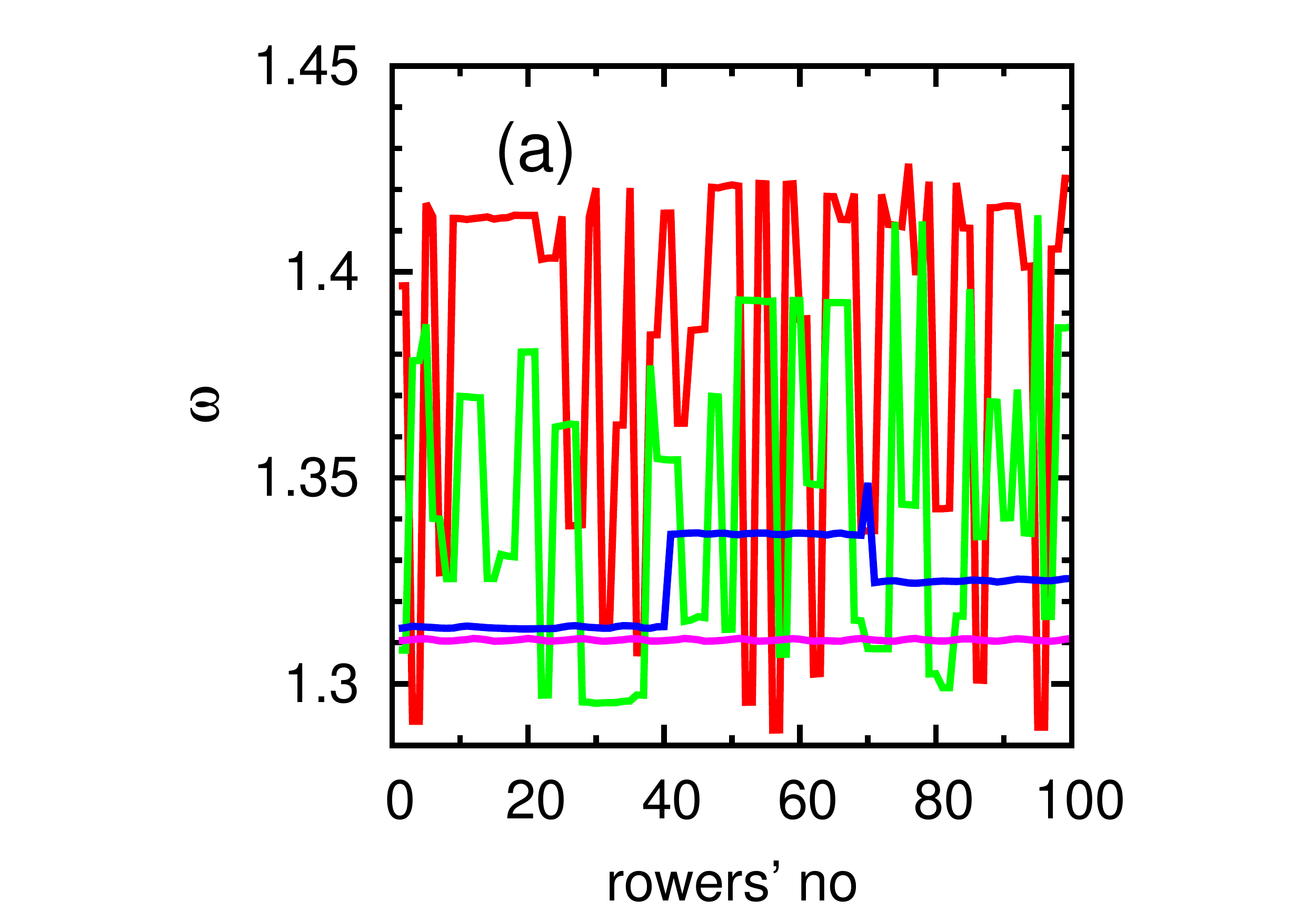}
\hspace{-1.9cm}
\includegraphics[scale = 0.23,angle=270]{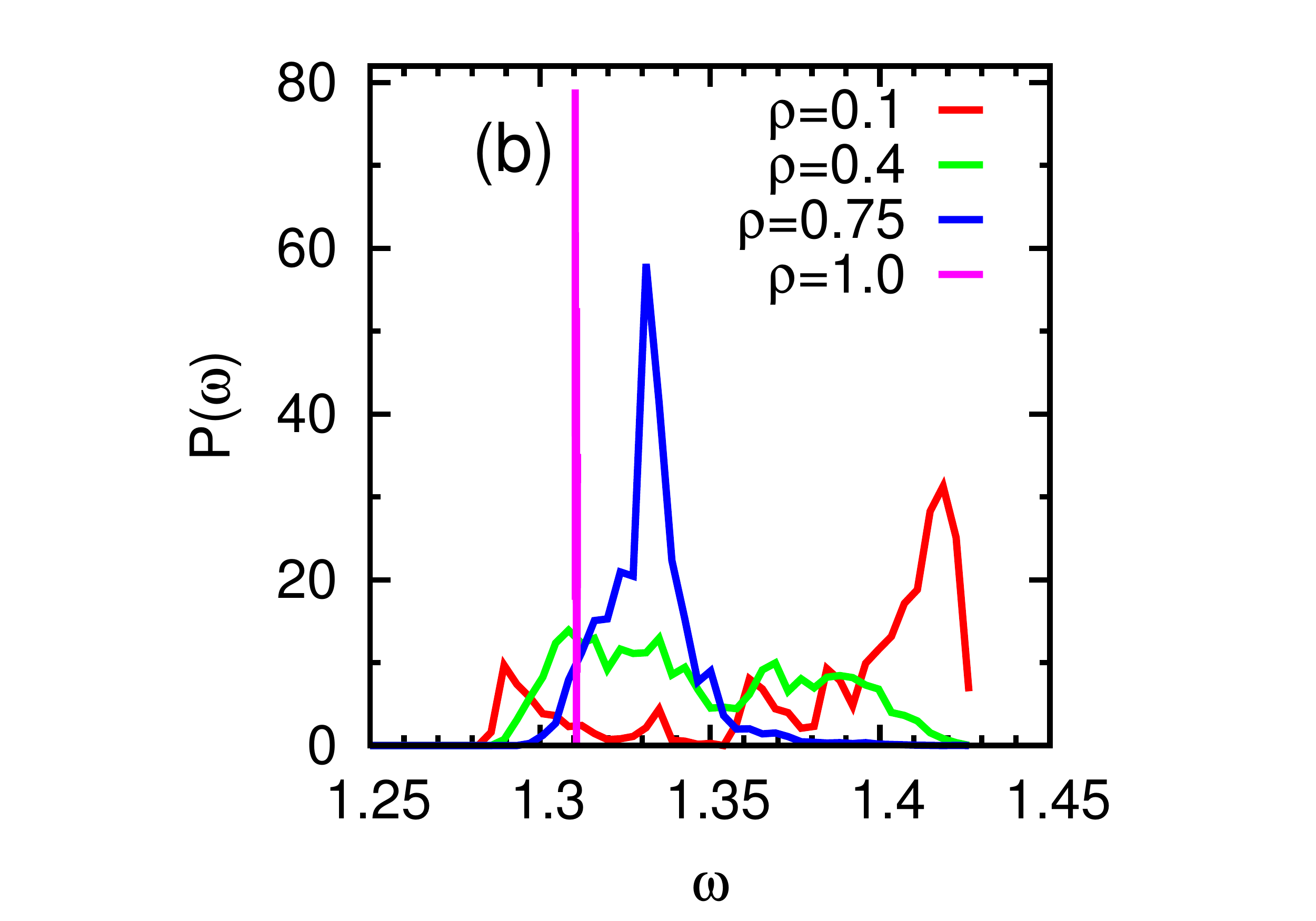}
}
\caption{(ii) Random case. Frequency plots for randomly positioned rowers with density $\rho=0.1, 0.4, 0.75$, and for a regular lattice with $\rho=1$. (a) The frequency of the rowers, $\omega$ is plotted as a function of rower's number. For $\rho=1$, all the rowers oscillates in a single collective frequency. For $\rho<1$, we see the whole system breaks into different parts, and these different local groups (synchronized island) oscillate with different frequencies. The number of such groups increases with decreasing $\rho$. (b) The probability distributions $P(\omega)$ are plotted for the same values of $\rho$ (as in (a)). For $\rho=1$,  it is a $\delta$-function (data is scaled by a arbitrary number for best visualization). As  $\rho$ is decreased, the width of the distribution increases, while the average $\omega$ increases. For $\rho=0.1,0.4$ multiple peaks are observed.}
\label{fig:fig13}
\end{figure}

In \fref{fig:fig13} (a), we have plotted the steady state frequency profile as a function of rowers' number for different densities in the random case. 
For a given $\rho$, we consider a configuration of randomly placed rowers on a lattice and evolve this system with a random initial condition of rowers.
 For $\rho=1$ (perfect regular array), all the rowers oscillates with the same frequency. This results in a $\delta$-function for $P({\omega})$  (\fref{fig:fig13} (b)). For $\rho<1$, various clusters of rowers start to oscillate at different frequencies, leading to a finite width in  $P({\omega})$ plot (\fref{fig:fig13} (b).   
Let us call {\it synchronized island} a group of consecutive rowers (connected by next occupied lattice sites) beating with the same frequency. We find that a synchronized island consists of a few clusters ($\sim$ 2-3, see Appendix~III). The number of such synchronized islands increases with decreasing $\rho$ (except for very small density where most rowers beat at their natural frequency) and so does the width of the distribution $P(\omega)$. From our study on finite numbers of rowers, we know that if the separation between two groups is less than a critical distance they oscillate with the same frequencies. The value of the frequencies depends on the separation between two groups and number of rowers in each group. This leads to a non-trivial distribution of the beating frequencies. 
 
For small densities, the frequency spectrum will be dominated by small clusters characteristic frequencies. Indeed,
for $\rho=0.1$, the position of the distant peak at $\omega \simeq \omega_0$ is due to isolated single rowers, while the other peaks at small $\omega$ are due to a collective effect of the rowers. For larger densities, the frequency spectrum is getting even more complex and broad.

\subsection{3. Case $(iii)$: Random clustered}

\begin{figure}[h]
\centering
\mbox{
\hspace{-1cm}
\includegraphics[scale = 0.23,angle=270]{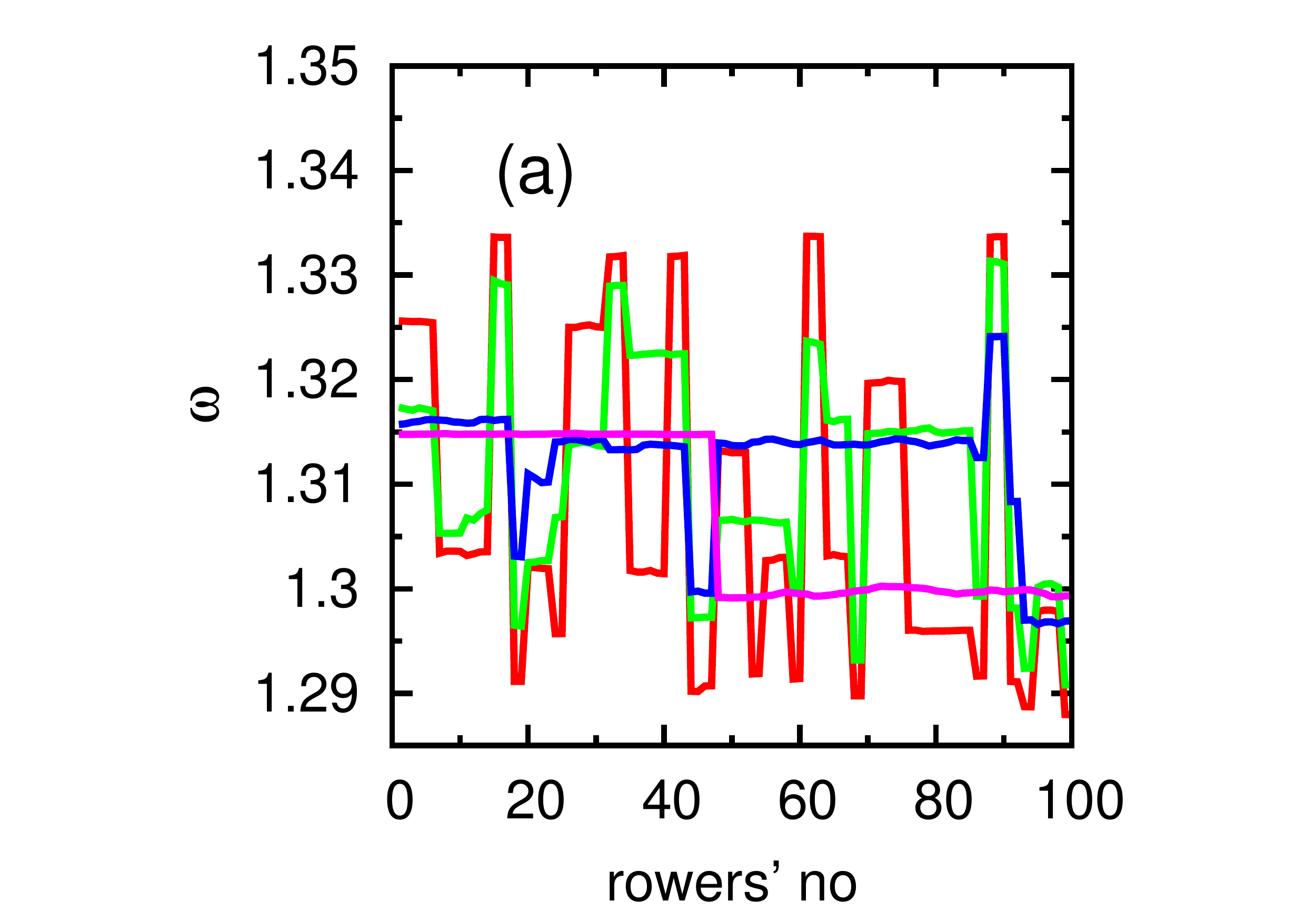}
\hspace{-1.9cm}
\includegraphics[scale = 0.23,angle=270]{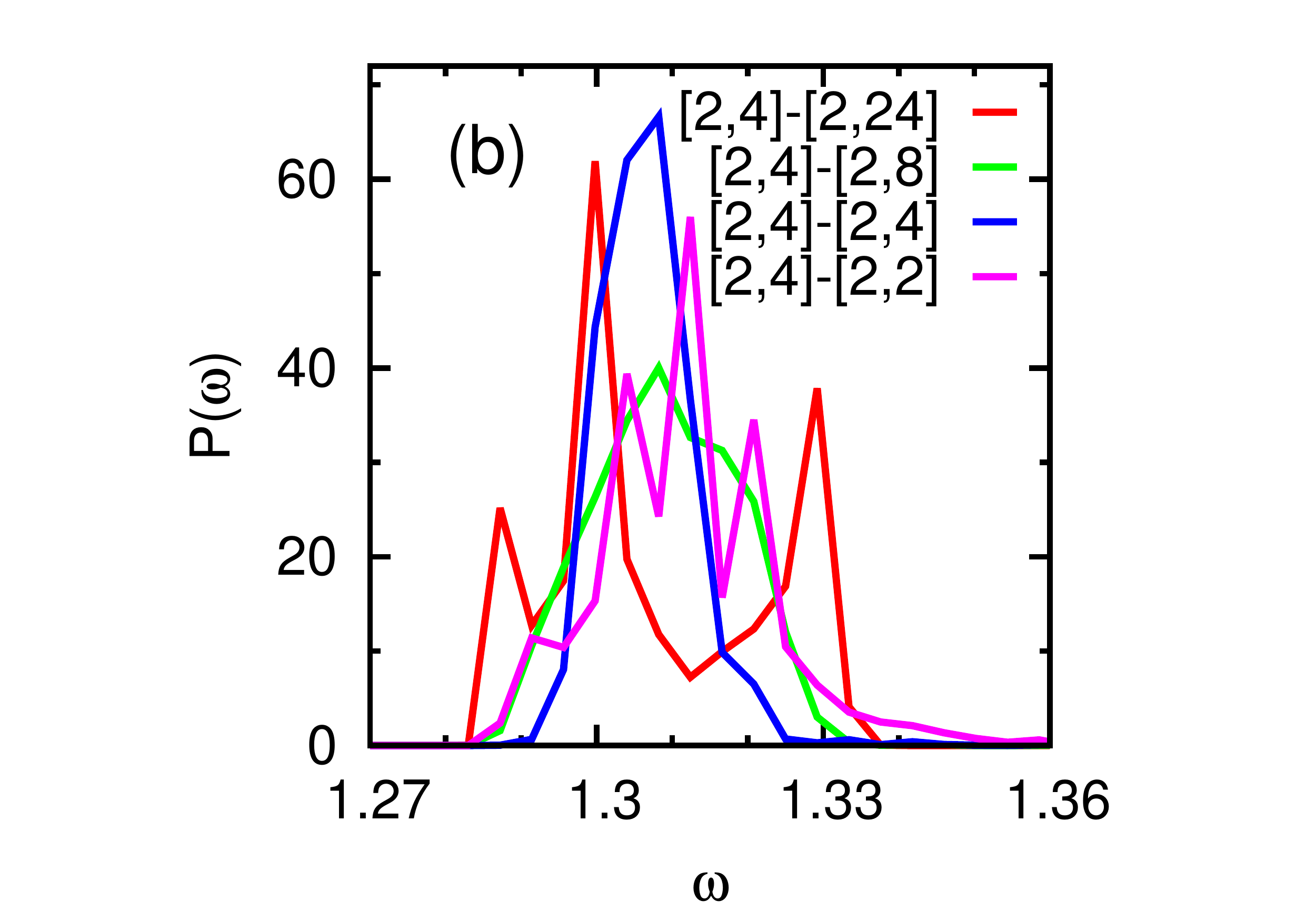}
}

\caption{(iii)  Random clustered case.  Frequency plots for ``[2,4]-[2,$gl_{max}$]''  lattices for $gl_{max}=2,4,8,$ and $24$.
The densities corresponding to these $gl_{max}$ are 
given by $\rho=$ 0.75, 0.6, 0.43, and 0.2 respectively. 
(a) The frequency of the rowers, $\omega$ is plotted as a function of rower's number 
for a given spatial structure of rowers' position. The whole system breaks into 
synchronized islands, which  oscillate with 
different frequencies. The number of such groups increases with 
increasing $gl_{max}$ (i.e. with decreasing $\rho$). 
(b) The probability distributions $P(\omega)$ are plotted for the same values of $gl_{max}$
as in (a). Remarkably, the mean and the width do not strongly depend on 
 $gl_{max}$.}
\label{fig:fig14}
\end{figure}

We consider [2,4]-[2,$gl_{max}$] random clustered heterogeneous configurations. The frequency profile and $P(\omega)$ are plotted for $gl_{max}=$2, 4, 6, 8  and 24 in \fref{fig:fig14}. As in the case of random heterogeneity, we observe that various clusters of rowers oscillate with different frequencies and the number of such clusters increases with  $gl_{max}$. This leads to finite widths in $P(\omega)$ (see \fref{fig:fig14} (b)). A remarkable feature is that the mean of the distribution is almost independent of the density  $\rho(gl_{max})$. For [2,4]-[2,$gl_{max}$] configurations such that  $gl_{max}>>1$, the average gap between two consecutive clusters is large and the effect of hydrodynamic couplings between them will be negligible in many cases. Indeed the peaks in $P(\omega)$ correspond to the collective frequencies of those clusters. The study with  [3,5]-[2,$gl_{max}$] configurations also leads to the same qualitative results (Fig. 15). We also find that a synchronized island consists of several numbers of clusters. This number increases as a function of the density of rowers, and for large densities, this number is very high compared to the random heterogeneity case (see Appendix~III).

\subsection{4. Comparing different structures with same density}

In the last sections, we have considered each type of heterogeneity separately and have investigated the nature of $P(\omega)$ for various densities. As experimental samples can  present a variety of spatial heterogeneities, we now compare the results for different heterogeneities for a given density. Moreover, many studies only report experimental values of $\rho$ and not the precise spatial distribution of cilia. Hence comparing theoretically the dynamical behavior of different configurations of cilia with fixed density will provide different types of scenarios that we hope we can compare to experimental observations and check whether this is consistent with the observed spatial arrangement of cilia.

\subsubsection{1. Frequency spectrum for different structures}

\begin{figure}[h]
\centering
\mbox{
\hspace{-1cm}
\includegraphics[scale = 0.23,angle=270]{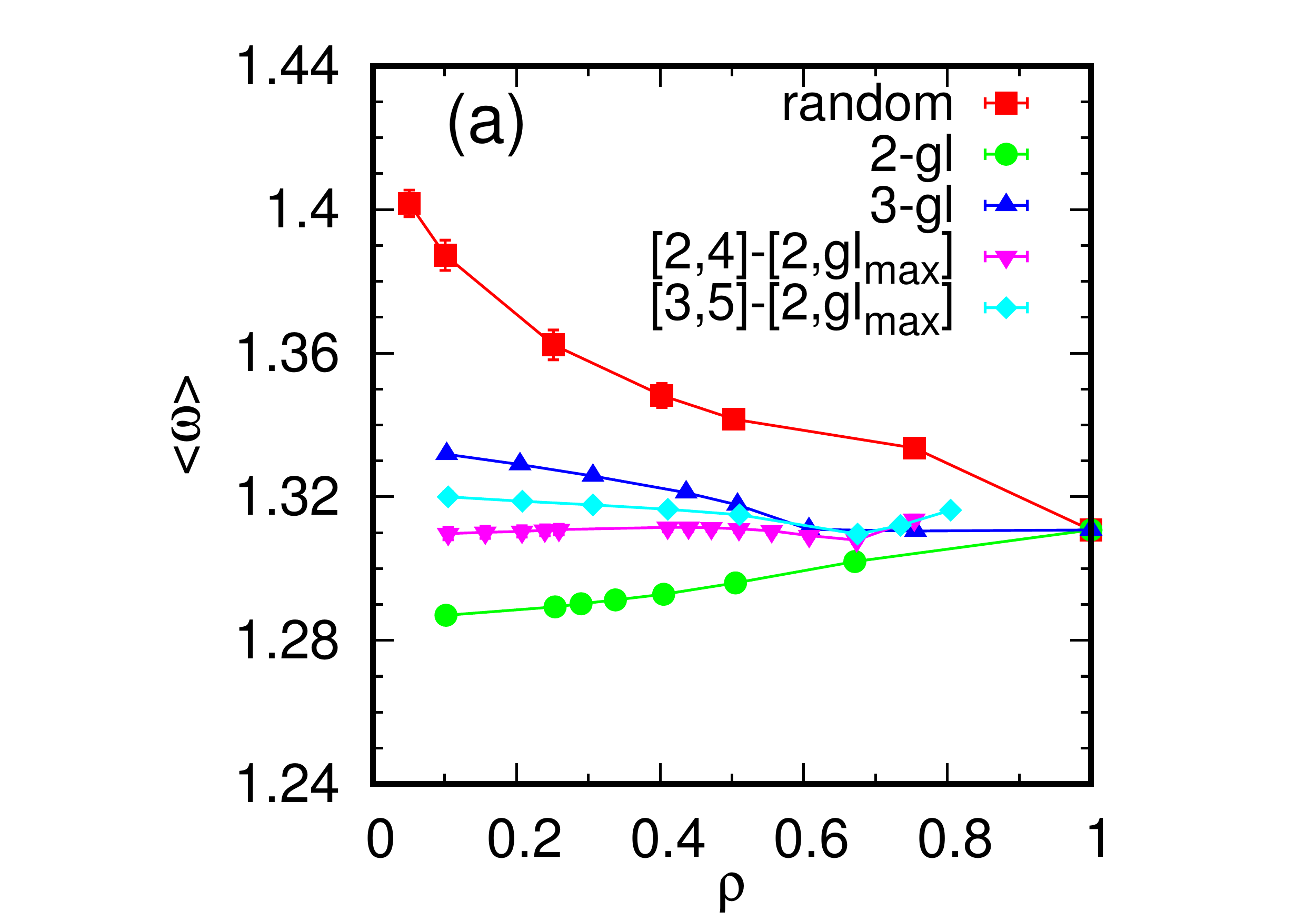}
\hspace{-1.7cm}
\includegraphics[scale = 0.23,angle=270]{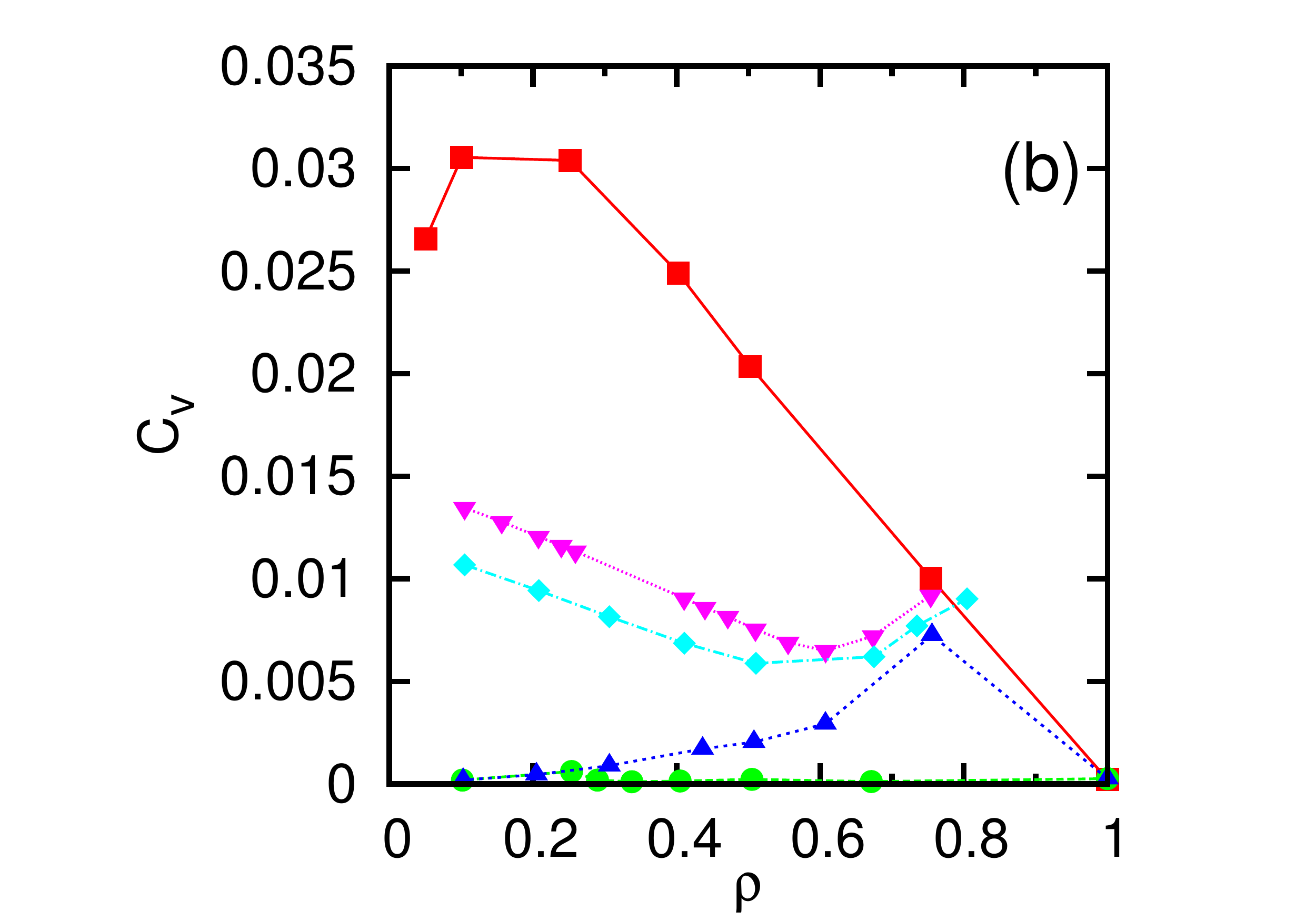}
}
\caption{Mean frequency and its fluctuation as a function of density $\rho$ for $\alpha=0.1$  and $N=100$. 
For random and random clustered lattice, the data are averaged over $100$ spatial configurations. 
For regular clustered lattice, the data  are averaged over $100$ initial configuration of rowers. For these 
plots we consider ``2-x'' and ``3-x'' regular clustered lattice, and  ``[2,4]-[2,x]' and ``[3,5]-[2,x]'' random clustered lattice.
(a) $\langle \omega \rangle$ against $\rho$ (with standard errors, errorbars are often smaller than point sizes), and (b) normalized variance $C_v$  versus $\rho$.  
}
\label{fig:fig15}
\end{figure}

We compare  the mean frequency $\la \omega \ra$ and the coefficient of variation $C_v= \sqrt{\langle \omega ^2 
\rangle -\langle \omega \rangle^2} / \langle \omega \rangle$
of the distributions $P(\omega)$. 
In \fref{fig:fig15}(a), we plot $\la \omega \ra$ as a function of density for different spatial heterogeneities 
of rowers. For random heterogeneity, $\langle\omega\rangle$ decays monotonically with density $\rho$. 
In the limit of $\rho \rightarrow 1$, all structures look like regular lattices and consequently all 
reach the collective frequency for many rowers. In the other limit $\rho \rightarrow 0$, $\langle\omega\rangle$  
converge to specific values depending on structures. In the case of random structures, $\langle\omega(\rho \rightarrow 0) 
\rangle = \omega_0$. In the case of regular clusters, $\langle\omega(\rho \rightarrow 0) \rangle = \omega_{coll,2}$ for ``2-x'' and 
$\langle\omega(\rho \rightarrow 0) \rangle = \omega_{coll,3}$ for ``3-x''.
In the case of random clusters, $\langle\omega (\rho \rightarrow 0) \rangle$ lies between $\omega_{coll,2}$  and $\omega_{coll,4}$ for 
``[2,4]-[2,x]'' and between $\omega_{coll,3}$  and $\omega_{coll,5}$ for ``[3,5]-[2,x]''. 
Note that $\la \omega \ra$ is almost independent of density in the random clusters case, in contrast with the results for other heterogeneity types. The standard deviation $C_v$, \fref{fig:fig15}(b), shows that the frequency distribution is the broader for random heterogeneity.

The average frequency $\omega$ appears to be a robust quantity for randomly clustered configurations of cilia, independently of the surface coverage $\rho$. 
This most probably corresponds to the type of surface coverage observed in in vivo samples. Consequently, this robustness implies that even at low densities ($\rho\simeq 0.1$ seems to be a commonly observed value), the collective frequency of cilia will be equal  to the frequency one would observe at any larger density. This sets a ``universal'' frequency for the cilia whatever the state of the surface coverage, given that it has the randomly clustered type of structure.

\subsubsection{2. Order parameter for different structures}

The results in the previous sections show that heterogeneity in
the rowers positions lead to partial loss of frequency locking.
Hence, all rowers do not oscillate with a common global 
frequency. But, some rowers do oscillate with a common 
frequency locally. This common local frequency is different 
in the different parts of the lattice and depends on the
local environment  of the rowers in a non-trivial manner.
The local frequency may also depend on the initial condition of 
the rowers. 

Hence, it is complementary to investigate whether there is any phase 
coherence among the rowers even in this partially frequency 
locked system. This might give an idea  about 
the local phase coherence of the rowers that oscillate
with the same frequency. In order to estimate phase coherence, we measure
the order parameter $A$ (defined in \eref{eqn:orderpara}) and 
compute its average value, $\langle A \rangle$
for different values of $\rho$ and different lattice structures, in the same spirit as what we did in the 3+1-rowers case in Section  II.5.
$\langle A \rangle$ is computed after  steady state time average  and ensemble average (average over several spatial 
configurations and/or several initial conditions of rowers). 

For $\rho=1$ the order parameter $A \simeq 0.9$, which is close but not equal to 1.
The definition of $A$ (\eref{eqn:orderpara}) shows that its value
depends on the distribution of phase difference of neighbours 
$\Delta\phi_j=\phi_{j+1}-\phi_{j}$. 
The phase locked angle is not 
perfectly $\pi$, rather we see a finite width of 
the distributions of $\Delta\phi_j$ which is (almost) independent of system sizes.
This leads a l value slightly less than $1$ (see Appendix \ref{app:orderpara} I).

Not surprisingly, the regular clustered configurations conserve the phase coherence, except at very low densities.
However, it is remarkable that the random clustered case is more coherent at all densities than the pure random case. In this case, which is likely to be the experimentally relevant case, spatially separated clusters are internally coherent, and all rowers beat with almost the same average frequency whatever its position and the density of the sample.

\begin{figure}[h]
\centering
\mbox{
\hspace{-1cm}
\hspace{-2cm}
\includegraphics[scale = 0.33,angle=270]{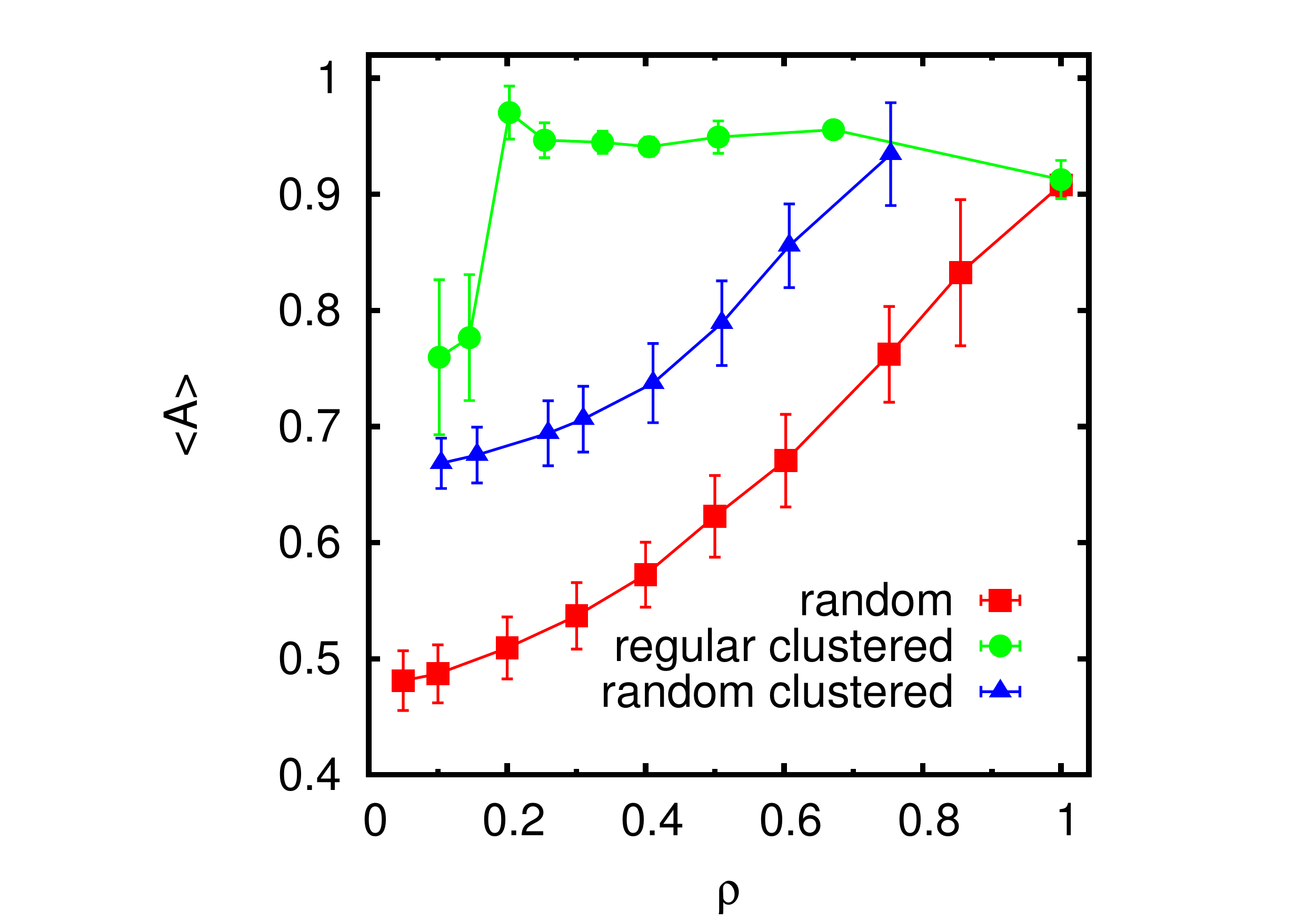}
}
\caption{Order parameter $<A>$ as a function of density $\rho$ for different structures for $N=100$. For random and random clustered lattice, the data are 
averaged over $100$ spatial configurations. For regular clustered lattice, the data  are averaged over $100$ initial configuration of rowers.  For these 
plots we consider ``2-x'' regular clustered lattice and  ``[2,4]-[2,x]'' random clustered lattice.} 
\label{fig:fig16}
\end{figure}

\section{V.  Conclusion}

While ciliated epithelium are most often modeled as large homogeneous carpets, they are observed experimentally to be inhomogenous and often rather sparse. 
However, cilia on such surfaces show coordination to a large extent, and thanks to this coordination fulfill their biological role. We have addressed this issue here, in a simplified model of a 1D heterogeneous array of rowers.

Hydrodynamic coupling is long-range and leads to collective coordination of the rowers phases in the homogeneous array previously studied.
When spatial heterogeneity is introduced,   gaps where rowers are absent lead to unperfected or lack of coordination. This is true in the simplest case of a very small number of rowers: if the gap length between two small clusters of rowers exceeds a critical distance, phase drifting is observed and ultimately leads to decoupling of rower clusters which are internally coordinated.
Interestingly this critical distance does not depend on the prefactor in the hydrodynamic interaction, which is proportional to the viscosity; it only depends on the internal parameters of the rower motility (but is likely to depend on the range of the interactions).

Looking at the coordination of two clusters of rowers as the spatial gap between them increases paves the way to understanding more complex arrays of rowers. We find that when the two clusters contain the same number of rowers, all the rowers beat at the same frequency whatever the gap. If they do not contain the same number of rowers, each cluster bifurcates to its own intrinsic characteristic frequency after the gap length exceeds a certain value. The distance after which the clusters are totally decorrelated (their frequency reach the one they would have in the absence of the other cluster) varies in a non trivial way with the size of the patches.

Hence the study of more complex arrays of rowers containing a large number of clusters of various sizes reveals a distribution of frequencies based on the physics described above. Not surprisingly, a regular configuration of clusters of similar size will show a delta distribution of frequencies. In contrast, a totally random configuration will produce a wider distribution of frequencies if the density of rowers is small, while the average frequency will decrease. In between those two cases, a randomly clustered configuration (consistent with experimental observations) has a distribution of frequencies that depend barely on the density of rowers, while the average beating frequency is close to a constant when the density is varied. Consequently a realistic carpet of cilia with density typically $\rho=0.1$ will have the same average frequency as a more dense carpet. Both the  frequency spectrum, and  the order parameter $A$ which quantifies the coordination of phase differences between neighboring rowers (existence of a metachronal wave) are complementary in providing a description of the dynamical states of the rowers.

While an experimental situation is obviously subject to thermal noise, our study and many others \cite{CosentinoPre03,StarkEpj11,CosentinoPRE12} are deterministic. In this approximations, the model presents a few complications as the dynamical states may marginally depend on initial conditions, as discussed in the text. This complication is removed as one would include noise, and we believe the results stay qualitatively the same \cite{CosentinoSM12}. Besides the thermal noise, there is an active source of noise which is intrinsic to the system. In the case of the flagellar beating for Chlamydomonas and sperm cells, it has been demonstrated that the later has a dominating contribution in their synchronized beating \cite{Goldstein2009, Ma2014}. It would be interesting to study the role of spatial heterogeneity in the presence of such intrinsic noise in future.

Moreover, other studies have considered a variation of hydrodynamical coupling, by taking into account the presence of a wall on which the rowers are attached (however without including noise \cite{Brumley2016,StarkEpj11} ). This variation introduces remarkable changes in the observed dynamical states of the cilia, like the coexistence of phase-locked and desynchronized clusters, known as chimera states. This is likely to happen in homogeneous arrays as studied in the above cited papers, and is likely to happen too in the heterogeneous case, though it has not been studied so far.

Along the same line, it will be interesting to study 2D carpets of rowers, as well as a realistic experimental spatially-resolved  sample. In this case, we believe that the orientation of beating of the rowers will also be responsible for the spatial correlations of the dynamical states  in 2D \cite{Dey_inprep}.

{\it Acknowledgments} - This work has been carried out thanks to the support of the LabEx NUMEV project ($\rm{n^o}$ ANR-10-LABX-20) and of the MUCOCIL project ($\rm{n^o}$ ANR-13-BSV5-0015) both funded by the Investissements d'Avenir French Government program, managed by the French National Research Agency (ANR). We thank our collaborators (D. Donnarumma, M. Jory, A. Bourdin, I. Vachier, A. Fort-Petit, D. Gras, P. Chanez, A. Viallat, K. Khelloufi) for fruitful exchange on their knowledge of bronchial epithelium. We also wish to thank P. Cicuta, L. M. Cosentino, and B. Friedrich for useful discussions on the model and results.



\setcounter{equation}{0}
\setcounter{table}{0}
\setcounter{section}{0}
\makeatletter
\renewcommand{\theequation}{A\arabic{equation}}
\renewcommand{\thetable}{A\arabic{table}}
\renewcommand{\bibnumfmt}[1]{[A#1]}
\renewcommand{\citenumfont}[1]{A#1}

\appendix*
\section{Appendix I. The order parameter A.}
\label{app:orderpara}

As we have discussed in the main text, the degree of synchronization in the system of phase oscillators 
can be measured by the following complex order parameter,
\bea
Z = A \, \rm{e}^{i \Phi} =  1/(N-1) \sum_{j=1}^{N-1} \rm{e}^{i \Delta\phi_j},  
\label{eqn:A1}
\eea
where $\Delta\phi_j=\phi_{j+1}-\phi_{j}$ are phase difference of nearest neighbours and $N$ is the number of the oscillators.  
The magnitude $A$ describes the phase-coherence, and polar angle $\Phi$ indicates average phase-difference of neighbours. 
The system show a maximal coherence when $A=1$, and no coherence for $A=0$. In the large $N$ limit, we can write the above 
summation by integration, 
\bea
A \, \rm{e}^{i\Phi} =  \int_{0}^{2 \pi}  d(\Delta\phi) \, P(\Delta\phi)  \, \rm{e}^{i \Delta\phi},
\label{eqn:A2} 
\eea
where $P(\Delta\phi)$ is the probability distribution of phase difference of nearest neighbours. 

Let us discuss three cases in detail for different types of phase coherence. 
{\bf Perfect phase locking :} In this case, the neighbouring pairs are phase locked to a constant angle $\delta$ i.e, $\Delta\phi_j=\delta$. 
Therefore, $P(\Delta\phi) = p(\delta - \Delta\phi)$. From \eref{eqn:A2} we get $A=1$, and we also see that $A$ is 
independent of $\delta$. {\bf No synchronization : } The probability distribution of $\Delta\phi$ is totally random. 
In this case, $P(\Delta\phi)=1/(2 \pi)$ which leads $A=0$. {\bf Partial phase locking :} The width (standard deviation) of 
the distribution of $\Delta\phi$ is neither zero (as in delta function) nor it possess a maximal width (as in uniform distribution). 
It rather has a finite width in between two extreme cases. Let us assume such a distribution 
by a normal distribution with standard deviation $\sigma$,
\bea
 P(\Delta\phi) = \frac{1}{\sqrt{2 \pi \sigma^2}} \rm{e}^{-\phi^2/{2 \sigma^2}}.
\eea  
For the above distribution, we can derive the expression for $A$ using \eref{eqn:A2}. After calculating the
Gaussian integral we get,
\bea
A= \rm{e}^{-1/2\sigma^2}.
\eea
Please note that $A\rightarrow 1$ when $\sigma \rightarrow 0$, and  $A\rightarrow 0$ as $\sigma \rightarrow \infty$.

\section{Appendix II. Critical distance for various parameters}
\label{app:critical_distance}

In Table.~\ref{tab:tab1}, we present the value of critical distance between two clusters at which phase drift behavior appears for five different sets of internal parameters of cilia $(k,s)$ for $N=3$ and $4$ (asymmetric case). For $N=3$, we observe that the value of $d_c$ is depends on $k$ and $s$ but, independent of $\alpha$ and initial conditions. However, for $N=4$ the value of $d_c$ is quite sensitive to $\alpha$ and initial conditions.

\begin{table}
\begin{center}
\begin{tabular}{ |c|c|c|c|}
 \hline
 (k, s)  & $\alpha$ & $d_c$ for $N=3$ & $d_c$ for N=4\\
 \hline
 (1.0, 0.8) &   0.01  & 2 & (3-5) 3.51 $\pm$ 0.02 \\
            &   0.05  & 2 & (3-4) 3.51 $\pm$ 0.01\\
            &   0.1   & 2 & (3-4) 3.51 $\pm$ 0.01 \\

\hline
(1.0, 0.95) &   0.01   & 4 &  (11-12) 11.51 $\pm$ 0.01\\
            &   0.05   & 4 &  (8-9) 8.50 $\pm$ 0.01\\
            &   0.1    & 4 &  (6-7) 6.44 $\pm$ 0.01\\

\hline
(1.1, 0.75) &   0.01   & 2 & (4-5) 4.49 $\pm$ 0.01\\
          &    0.05  & 2 & (3-4) 3.51 $\pm$ 0.01\\
          &    0.1   & 2 & (3-4) 3.51 $\pm$ 0.01\\

\hline
(1.1, 0.8) &   0.01   & 3 &  (5-7) 5.52 $\pm$ 0.02\\
          &    0.05   & 3 &  (5-6) 5.49 $\pm$ 0.01\\
          &    0.1    & 3 &  (4-5) 4.50 $\pm$ 0.01\\

\hline
(1.2, 0.7) &   0.01   & 2 & (4-5) 4.49 $\pm$ 0.01 \\
          &    0.05  & 2 & (4) 4\\
          &    0.1   & 2 & (3-4) 3.51 $\pm$0.01\\

  \hline
\end{tabular}
\caption{The critical distance $d_c$ for five different sets of internal parameters of cilia $(k,s)$ for $N=3$ (column 3) and $4$ (asymmetric case, column 4). For each set of $(k,s)$, we consider three different values of hydrodynamic coupling strength $\alpha=0.1$, $0.05$, and $0.01$. For the computation of each $d_c$, $1000$ initial configurations are used. The integration step $h= 10^{-3}$. As the value of the lattice constant in our computation is $1$, the value of $d_c$ is an integer.  For $N=4$, $d_c$ is sensitive to initial configuration of rowers and its value lies between $(d_{\rm max}-d_{\rm min})$, as shown in column 4. The average values with error are also presented.}
\label{tab:tab1}
\end{center}
\end{table}

\section{Appendix III. Synchronized island}

\begin{figure}[!htbp]
\mbox{
\hspace{-1cm}
\includegraphics[angle=-90,scale=0.245]{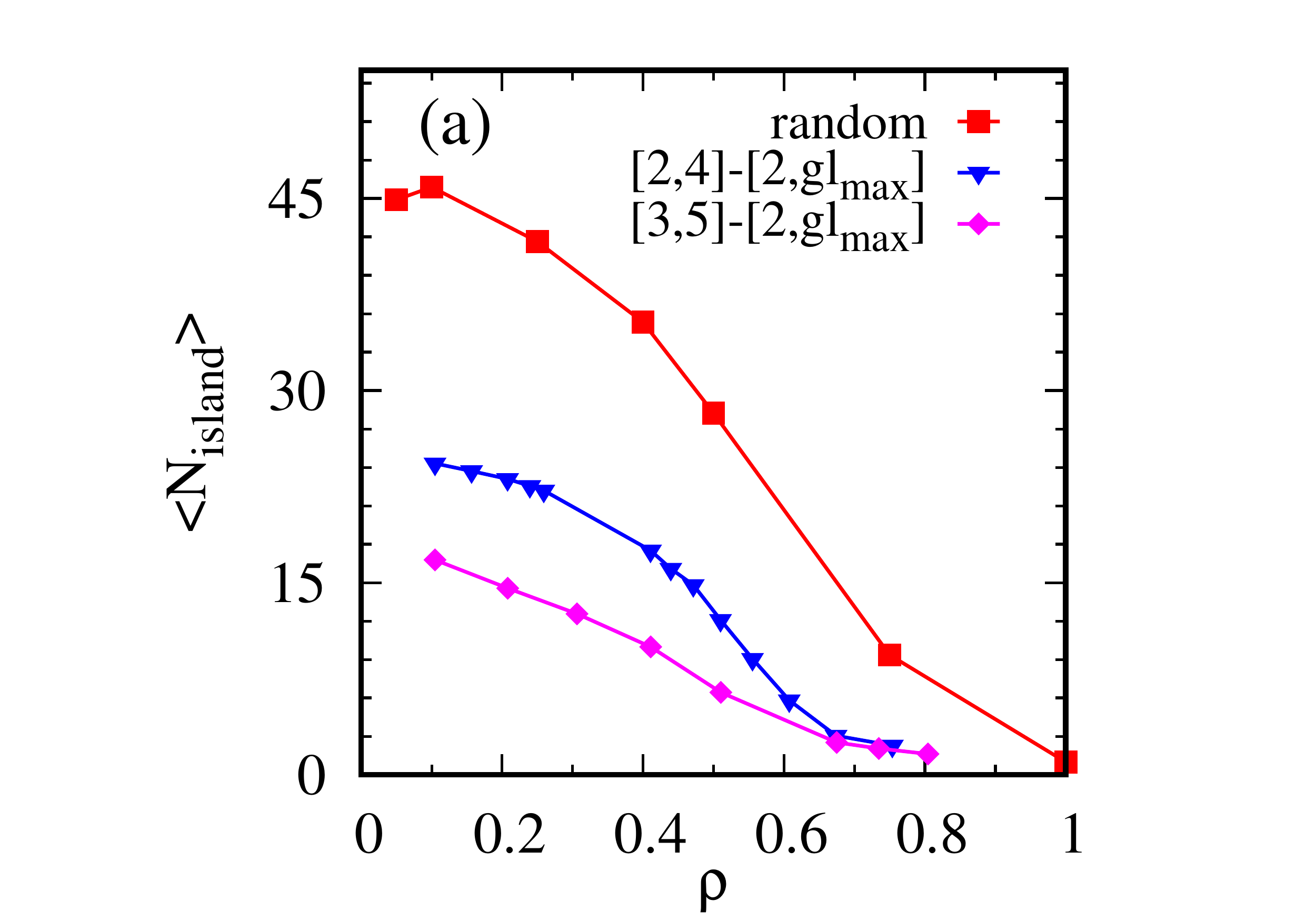}
\hspace{-2cm}
\includegraphics[angle=-90,scale=0.245]{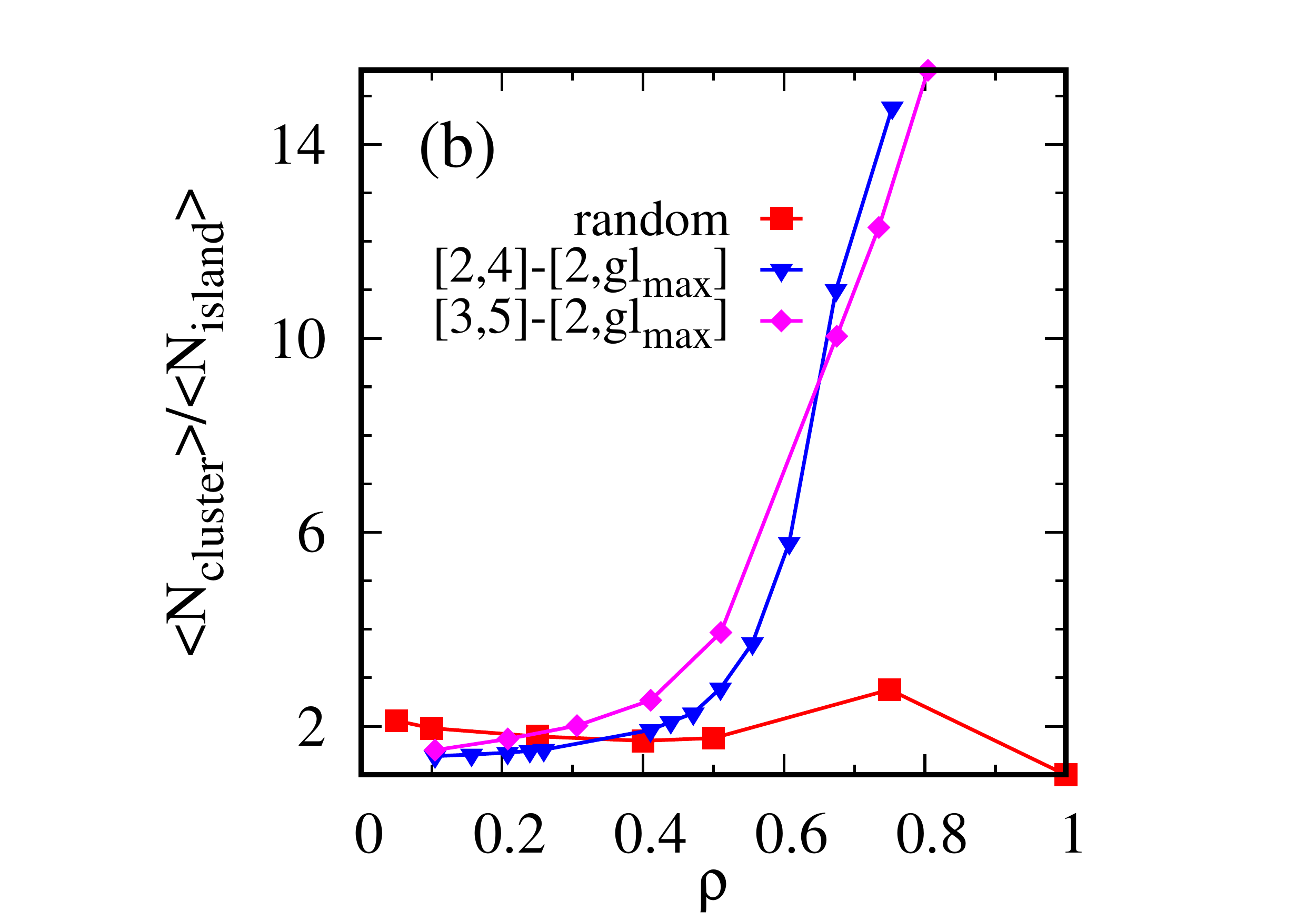}
}
\caption{Plots of synchronized island for different heterogeneities. (a) $\langle N_{cluster} \rangle$ is plotted against $\rho$. (b) $\langle N_{cluster} \rangle/\langle N_{island} \rangle$ against $\rho$.}
\label{fig:figs_synchoisland}
\end{figure}

Let us recall the definition of a synchronized island: a group of consecutive rowers (connected by next occupied lattice sites) beating in a common frequency. We have measured this quantity for random and random clustered heterogeneity for the same data provided in Fig.~\ref{fig:fig13} and Fig.~\ref{fig:fig14} in the main text. In Fig.~\ref{fig:figs_synchoisland}, we plot the average number synchronized islands ($\langle N_{cluster} \rangle$) and the ratio of the average number of clusters to the average number synchronized islands ($\langle N_{cluster} \rangle/\langle N_{island} \rangle$). We observe that for all heterogeneity $\langle N_{island} \rangle$ decreases with increasing $\rho$ and the value of $\langle N_{island} \rangle$ is large for random heterogeneity compared to random clustered heterogeneity, suggesting that the synchronization is more vulnerable to random heterogeneity. From Fig.~\ref{fig:figs_synchoisland} (b), it is clear that a synchronized island consists of several clusters. This number increases with decreasing $\rho$.

\label{app:synchronized_island}

\begin{figure}[]
\mbox{
\includegraphics[angle=-90,scale=0.25]{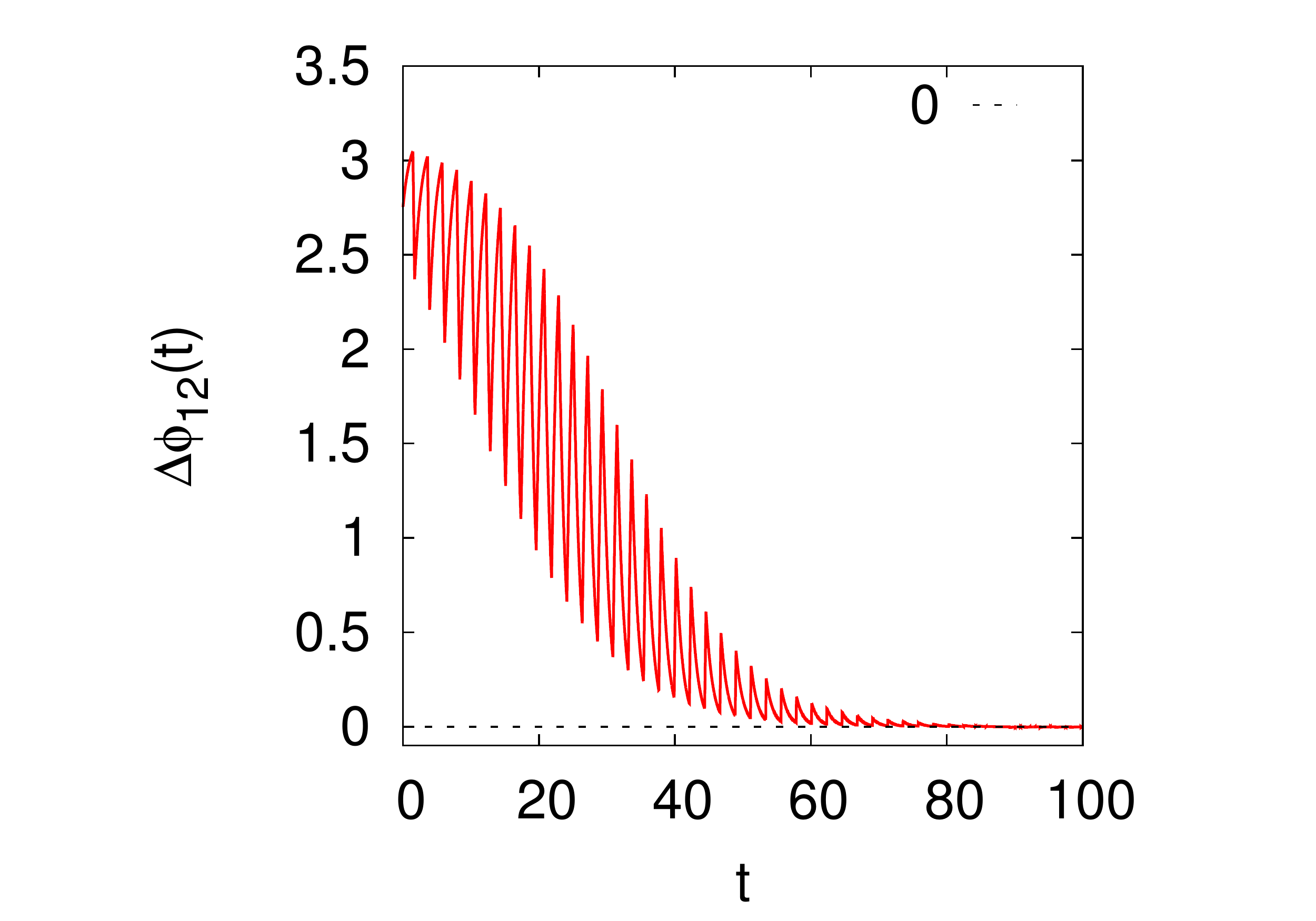}
}
\caption{In phase synchronization for a two rower system. The phase difference $\Delta\phi_{12}=\phi_{2}-\phi_{1}$ is plotted as a function of time $t$. 
At $t=0$ the rowers start from an arbitrary initial condition. The phase difference vanishes as $t$ increases. 
}
\label{fig:figs1}
\end{figure}

\begin{figure}[]
\centering
\mbox{
\hspace{-1cm}
\includegraphics[scale = 0.243,angle=270]{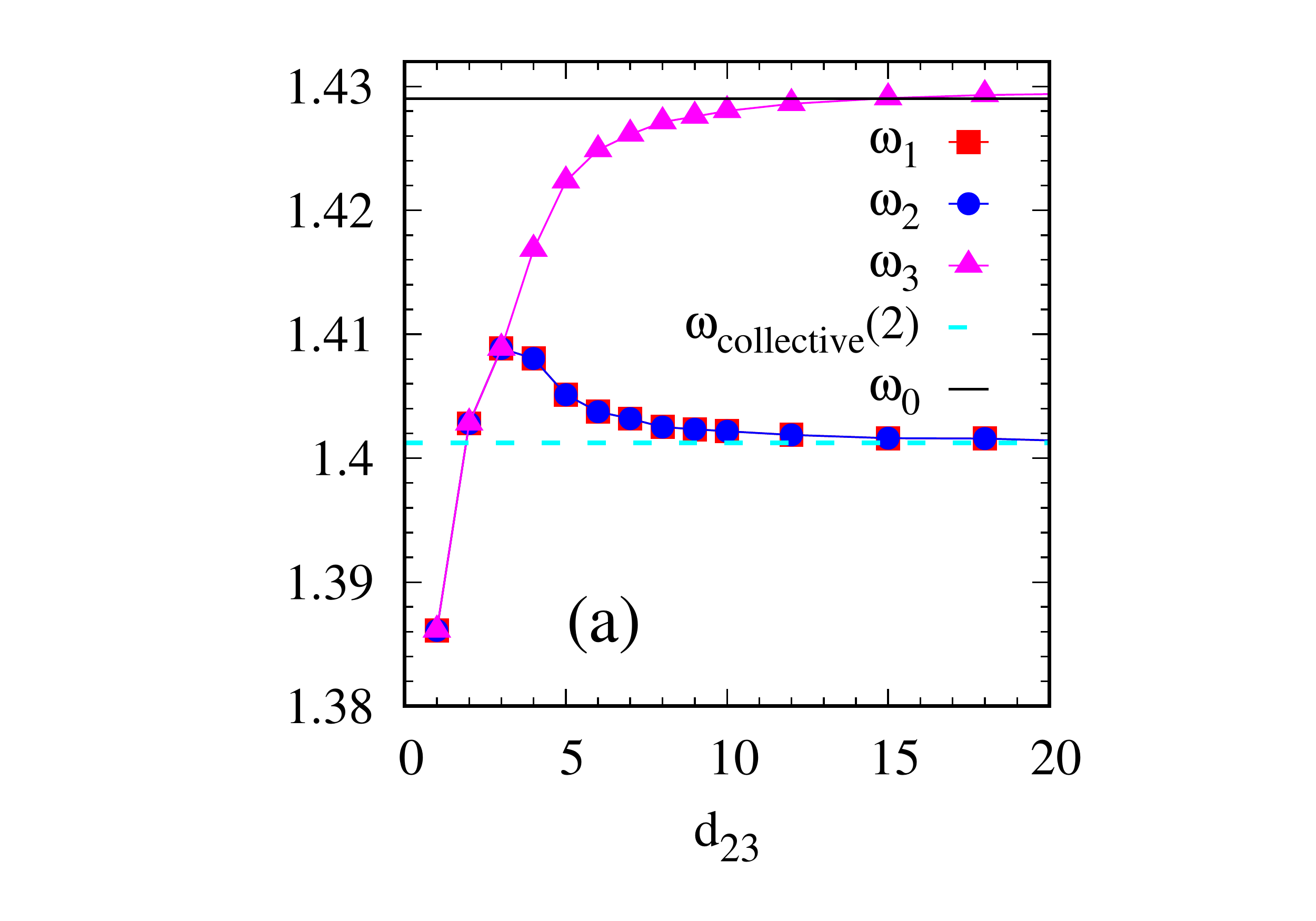}
\hspace{-2cm}
\includegraphics[scale = 0.243,angle=270]{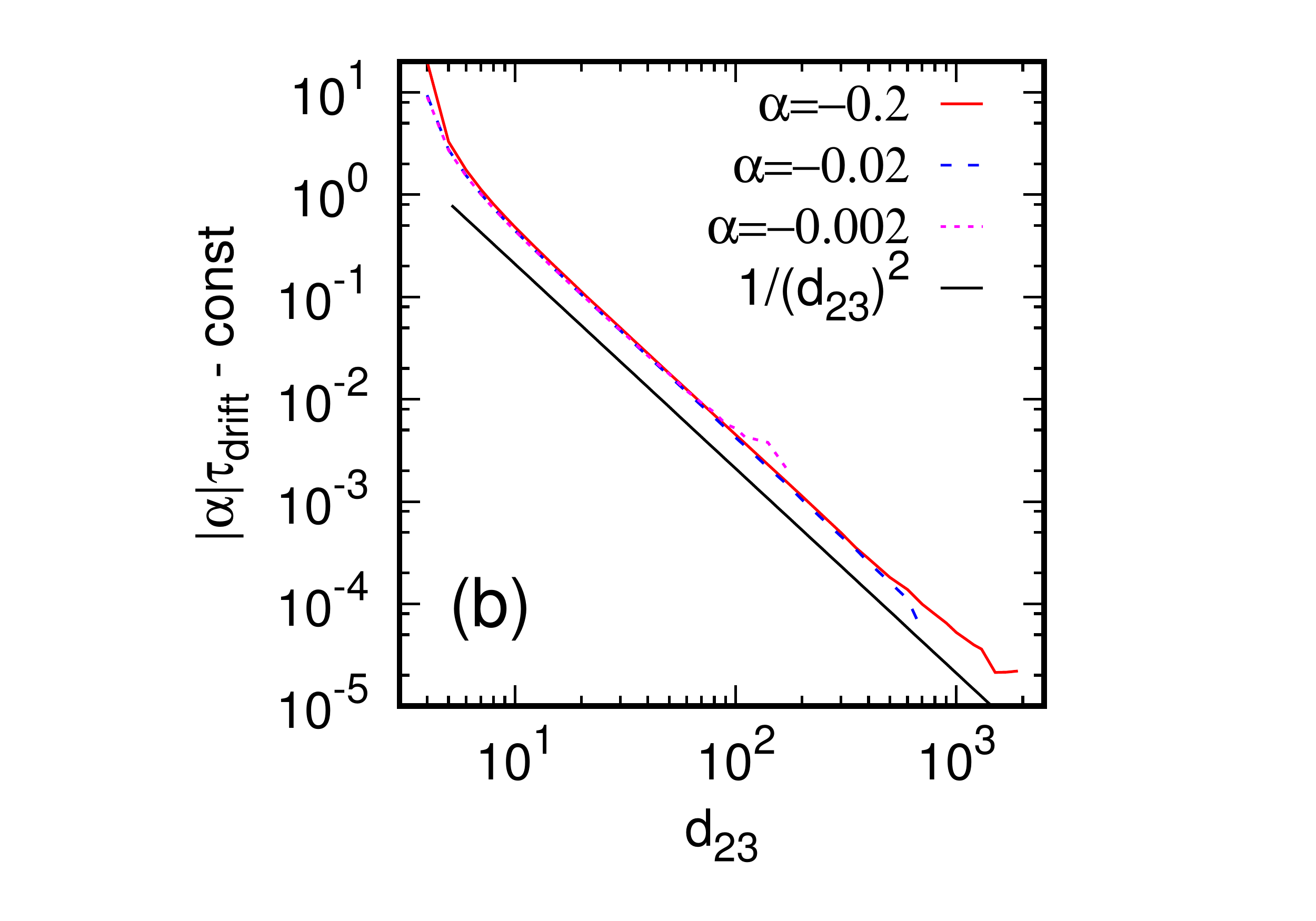}
}
\caption{(a) Angular frequencies of three rowers as a function of $d_{23}$ is plotted for $\alpha=-0.02$. Bifurcation occurs 
for $d_{23}\ge 4$. For very large $d_{23}$, rower 3 become almost independent of others and oscillates with its natural 
frequency $\omega_0$ while first two rowers oscillates with the collective frequency of two rower system. 
(b) The log-log plot of $\tau_{\rm drift}$ for $\alpha=-0.2, -0.02$ and $-0.002$. We multiply the data with respective $|\alpha|$. 
This scaling makes all the curves fall into a single curve $g(d_{23})$.  The 
function $g(d_{23})$ shows $1/d^2_{23}$ decay.
} 
\label{fig:figs2}
\end{figure}

\begin{figure}[]
\centering
\mbox{
\includegraphics[scale = 0.3,angle=270]{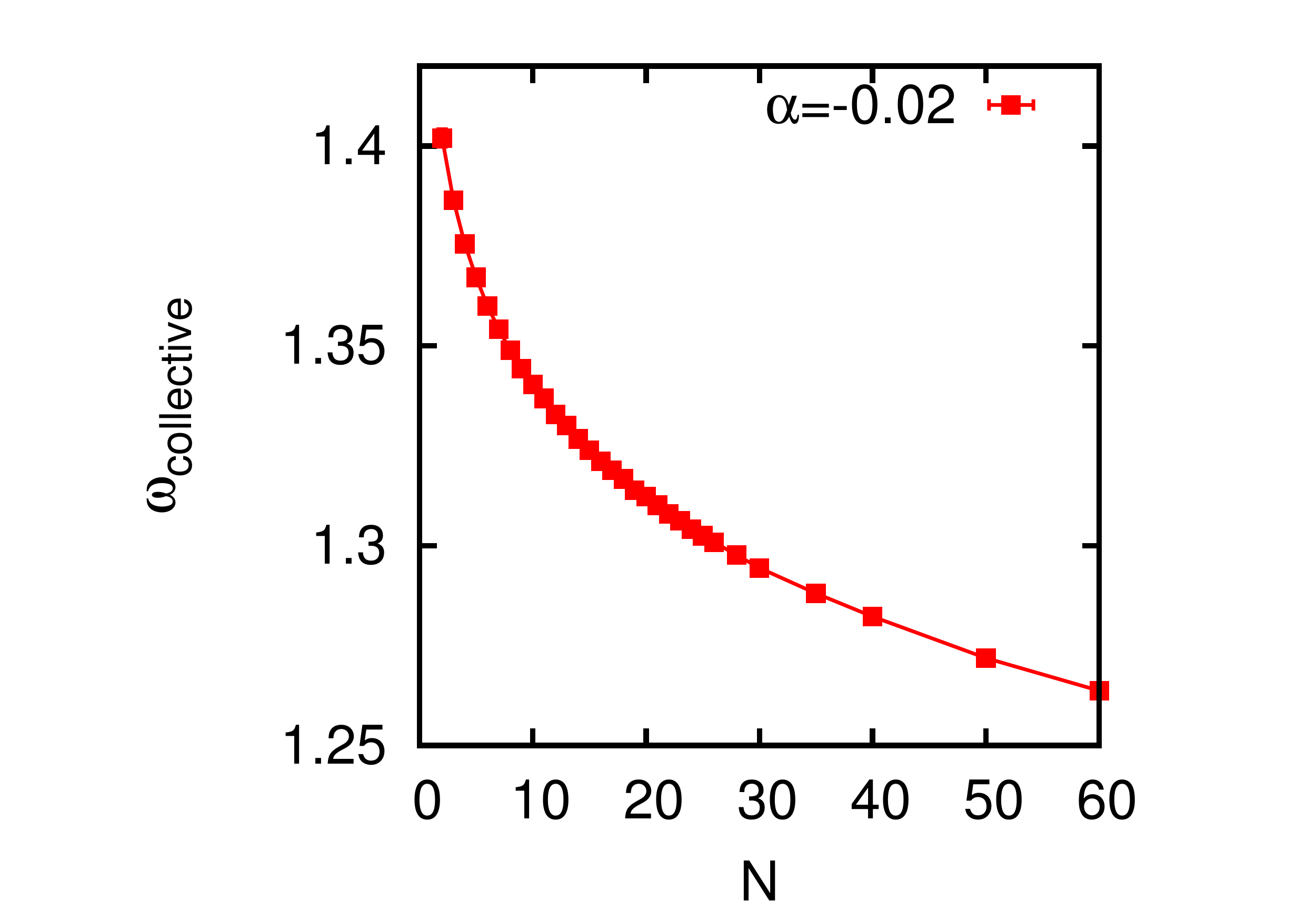}
}
\caption{The collective frequency is plotted as a function of total number of rowers $N$. 
The rowers are placed on a regular lattice with lattice constant $d=1$. The data are averaged over 
$1000$ initial configurations. It decays with system size $N$.}
\label{fig:figs3}
\end{figure}

\section {Appendix IV. The attractive case $\alpha<0$.}

In the main text, we have discussed the effect of spatial heterogeneity in the case when two rowers oscillate in anti-phase, and many rowers oscillate collectively as metachronal waves.  Here, we discuss the effect of spatial heterogeneity, in the case when rowers show in-phase synchronization. While for a system of rowers, the anti-phase solution can be achieved using a realistic hydrodynamic interaction ($\alpha>0$) (as discussed in the main text), the in-phase synchronization can be realized through a ``non-realistic''  hydrodynamic coupling with negative $\alpha$ \cite{CosentinoPre03} or using a negative force constant $k$ of harmonic driving force of rowers \cite{StarkEpj11}. We investigate the case of in-phase synchronization using the same dynamical equation \eref{eqn:dynamics} but with negative $\alpha$. Here, we present the results for $\alpha=-0.02$ and the same values of the force constant $k$ and amplitude $s$ as mentioned in the main text are used.

\subsection{Few rowers}

In \fref{fig:figs1} we show that two rowers, initially having arbitrary phases, oscillate in the same phase 
after a transient period.  For a three rower system, the phenomena of phase drifting has seen as the distance between 2nd 
and 3rd rowers $d_{23}$ is varied. In \fref{fig:figs2} (a), we plot the frequency of all the 3 rowers. 
Note that a bifurcation occurred for $d_{23}\geq d_c(=4)$.  In \fref{fig:figs2} (b), we show 
 the drift time $\tau_{\rm drift}$ for different values of negative $\alpha$. As in the case of positive $\alpha$, 
 the similar scaling of  $\tau_{\rm drift}$ holds for negative $\alpha$:  $|\alpha|\,\tau_{\rm drift}(\alpha, d_{23}) 
 = g(d_{23}) + const$. However, the decay of scaling function $g(d_{23})$ is different
from the scaling function $f(d_{23})$ for positive $\alpha$ (see main text).   Here, the scaling function  $g(d_{23})$ decays as $g(d_{23}) \sim 1/d^2_{23}$ which is  faster than the decay of $f(d_{23})$.

\subsection{Many rowers}

\begin{figure}[]
\centering
\mbox{
\hspace{-1cm}
\includegraphics[scale = 0.23,angle=270]{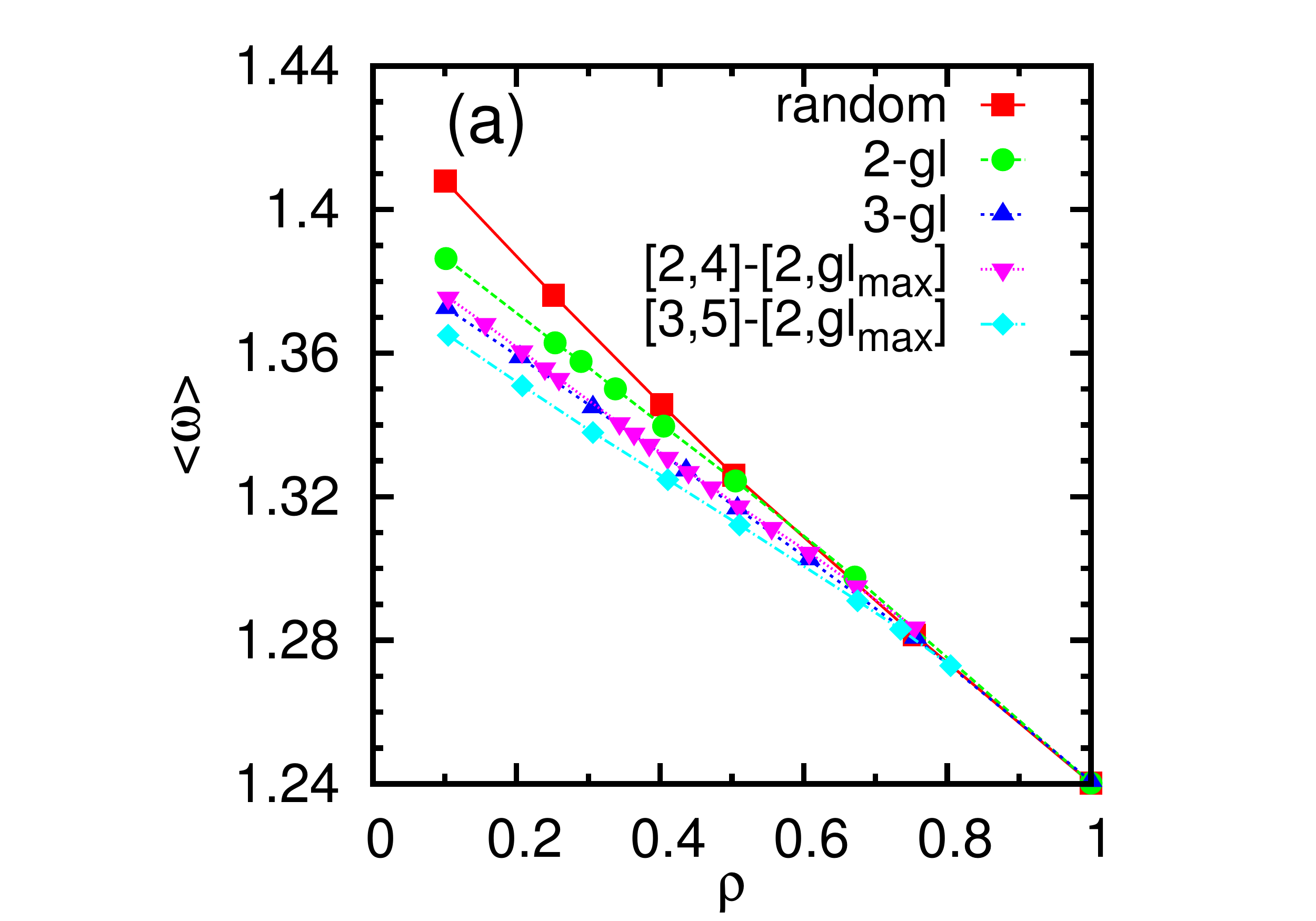}
\hspace{-1.5cm}
\includegraphics[scale = 0.23,angle=270]{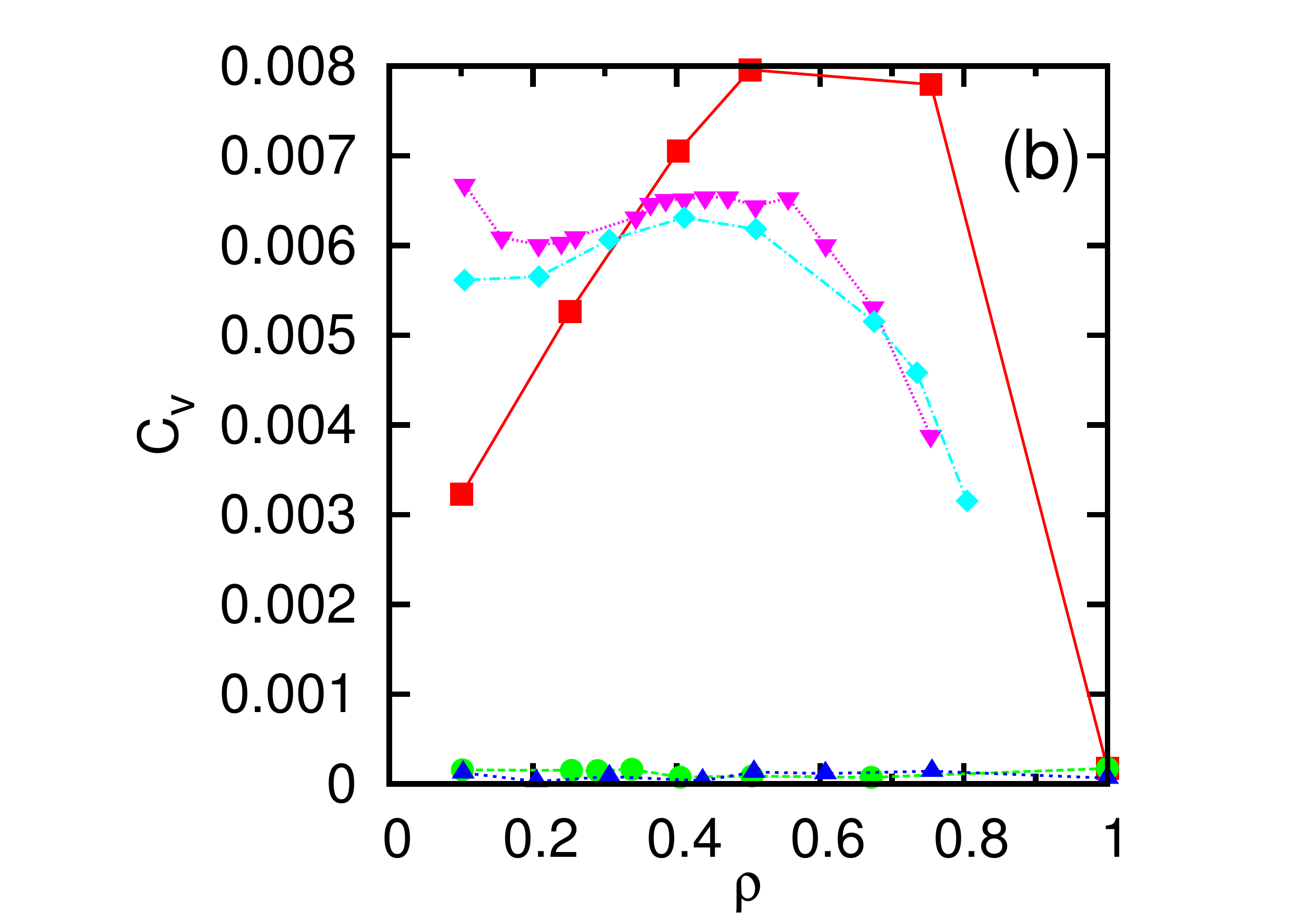}
}
\caption{Mean frequency and its fluctuation as a function of density $\rho$ for $N=100$. 
For random and random clustered heterogeneity, the data are averaged over $100$ spatial configurations. 
For regular clustered heterogeneity, the data  are averaged over $100$ initial configuration of rowers.  For these 
plots we consider ``2-x'' and ``3-x'' regular clustered, and  ``[2,4]-[2,x]' and ``[3,5]-[2,x]'' random clustered heterogeneity.
(a) $\langle \omega \rangle$ against $\rho$, and (b) the coefficient of variation $C_v$  against $\rho$.  
}
\label{fig:figs4}
\end{figure}

\begin{figure}[]
\centering
\includegraphics[scale = 0.3,angle=270]{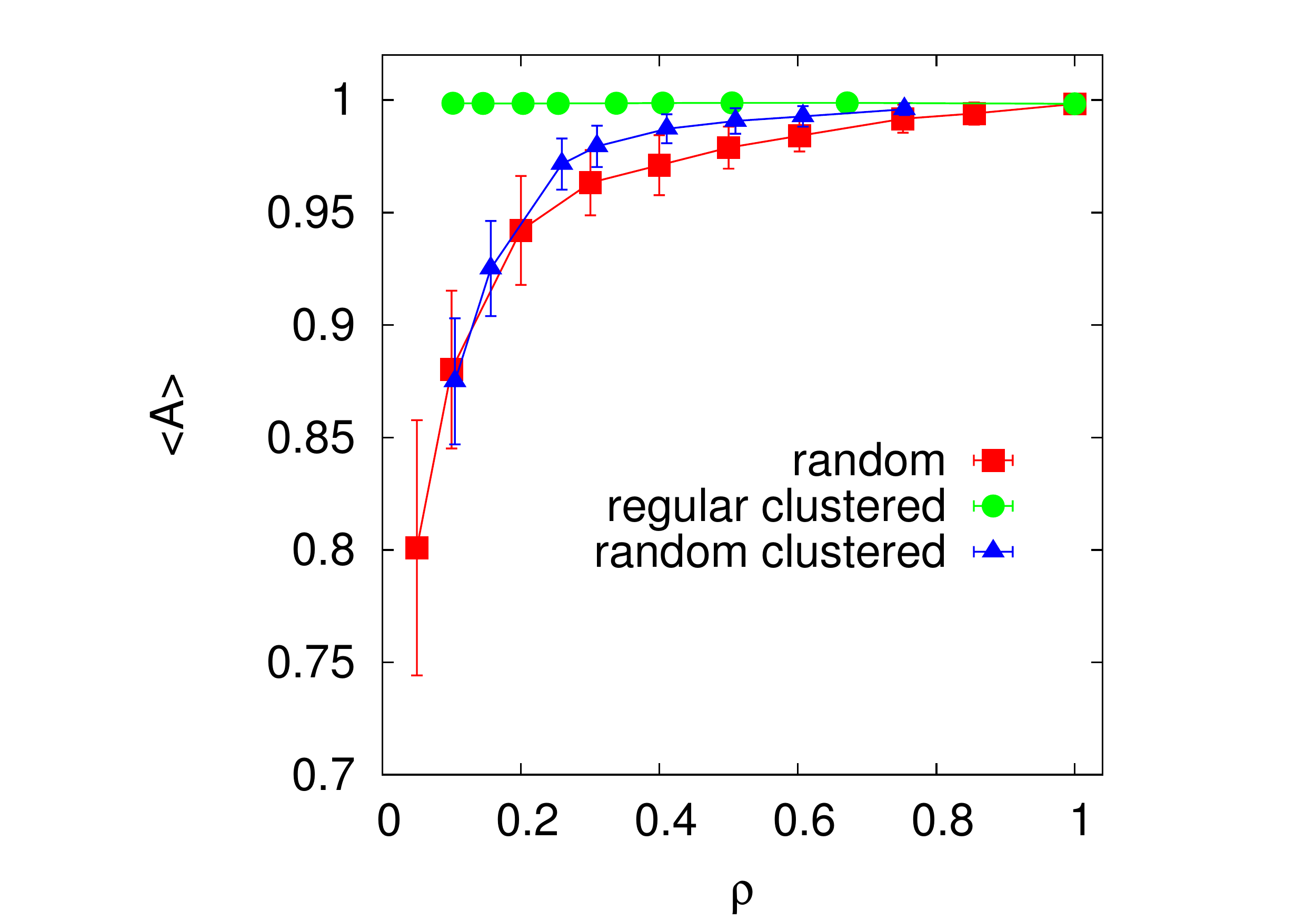}
\caption{Order parameters as a function of density $\rho$ for different structures for $N=100$. For random and random clustered lattice, the data are averaged over $100$ spatial configurations. For regular heterogeneity, the data  are averaged over $100$ initial configuration of rowers.  For these plots we consider ``2-x'' regular lattice and ``[2,4]-[2,x]'' random clustered lattice. For regular lattice, the phases of the rowers coherent for all the density. For random clustered and random lattice, the synchronization of phases gets destroyed at lower densities.
} 
\label{fig:figs5}
\end{figure}

We first consider $N$ rowers on a regular lattice with lattice constant $d=1$. In this case, all the rowers oscillate with a single common frequency $\omega_{coll,N}$. The phase of a rower and its adjacent neighbors are in-phase synchronized.  In \fref{fig:figs3}, we plot collective frequency as a function of system size $N$. We observe that $\omega_{coll,N}$ decays with $N$.

Next, we present the results of three different types of heterogeneous lattices (regular clustered, random and random clustered) for $\alpha<0$. We find that the spatial inhomogeneity in rowers' position leads to fluctuation in the beating frequency of the rowers and reduces the order in phase coherence. In \fref{fig:figs4}(a), we plot average frequency of the rowers. We observe that $\la \omega \ra$ decreases as a function of the density $\rho$ for all the types of heterogeneous lattices.  In \fref{fig:figs4}(b) we plot the coefficient of variation $C_v$ which is a normalize fluctuation of frequency.  We observe that for a fixed $\rho$, $C_v$ depends  on the type of heterogeneity. For regular clustered $C_v$ is less and for random heterogeneity $C_v$ is high, while for random clustered it is intermediate.

The phase coherence can be measured by the order parameter $A$ (defined in the main text). The order parameter $A$ is plotted as function of density $\rho$ in \fref{fig:figs5}.

\end{document}